\documentclass[english,10pt,aps,prd,a4paper,preprintnumbers,floatfix,nofootinbib,showpacs,superscriptaddress, notitlepage]{revtex4-2} 
 \pdfoutput=1
\usepackage[usenames,dvipsnames]{color}  
\usepackage{graphicx}
\usepackage{caption}
\usepackage{subcaption}
\usepackage{amsfonts}
\usepackage[export]{adjustbox} 
\usepackage{amsmath}
\usepackage{amssymb}
\usepackage{float}
\usepackage[colorlinks=true,citecolor=darkred,urlcolor=darkred, pdfborder={0 0 0}]{hyperref}
\usepackage[normalem]{ulem}
\usepackage{braket}

\makeatletter
\def\p@subsection{}
\makeatother

\definecolor{darkred}{rgb}{0.6,0,0}

\definecolor{linkcolor}{rgb}{0,0,0.5}

\def\gsim{\raise0.3ex\hbox{$\;>$\kern-0.75em\raise-1.1ex\hbox{$\sim\;$}}}
\def\lsim{\raise0.3ex\hbox{$\;<$\kern-0.75em\raise-1.1ex\hbox{$\sim\;$}}}

\def\beqn#1{\begin{equation}\label{#1}}
\def\eeqn{\end{equation}}

\def\beqa#1{\begin{eqnarray}\label{#1}}
\def\eeqa{\end{eqnarray}}

\def\0nbb {$0\nu\beta\beta$ }

\def\Z2{$\mathcal{Z_2}$}

\newcommand {\ignore}[1]{}

\def\321{$\mathrm{SU(3) \otimes SU(2) \otimes U(1)}$ }

\newcommand{\AddrHBNI}{
	Homi Bhabha National Institute, BARC Training School Complex, Anushakti Nagar, Mumbai 400094, India }

\begin{document}

\bibliographystyle{unsrt} 

\title{Pseudo scalar dark matter in a generic U$(1)_X$ model}
\author{Shivam Gola}
\email{shivamg@imsc.res.in}
\affiliation{The Institute of Mathematical Sciences,
C.I.T Campus, Taramani, Chennai 600 113, India}
\affiliation{\AddrHBNI}

\preprint{EPHOU-21-017}

\begin{abstract}

We consider a $U(1)_X$ extension of the Standard Model~(SM), where the spontaneous breaking of $U(1)_X$ gauge group results in a pseudo scalar particle which is the proposed candidate for dark matter. In the model, we introduce three right-handed neutrinos~(RHNs) $N_R^i$ and two extra scalars $\Phi$, $\chi$, which are SM gauge singlets but charged under $U(1)_X$ gauge group. Right-handed neutrinos are required to have the model anomaly free and explain the neutrino oscillation data. The heaviest scalar breaks the $U(1)_X$ gauge symmetry and the other extra scalar gives a pseudo scalar DM candidate. A pseudo scalar dark matter~(DM) is an interesting candidate as it naturally evades the stringent direct detection bounds due to its coupling structure. We study the phenomenology of this pseudo-scalar DM while considering several theoretical and experimental constraints. We find that in our model, there is a feasible parameter space, which satisfies by the DM lifetime bound, relic and direct detection constraints while respecting the colliders and other bounds.

\end{abstract}

\maketitle
\section{Introduction}

Dark matter~(DM) is required to explain the missing mass problem from the observations of several astrophysical and cosmological objects such as galaxy rotation curves, gravitational lensing, Bullet clusters and cosmic microwave background~\cite{Garrett_2011,Profumo:2009tb,Bertone:2004pz,Bartelmann:1999yn,Clowe:2003tk,Harvey:2015hha,Hinshaw:2012aka,Ade:2015xua} etc. However, the nature of dark matter is largely unknown. It is known that none of the Standard model~(SM) particles can be a candidate for dark matter. Therefore, one must look for physics beyond Standard model~(BSM). Several particle candidates for DM have been proposed in the literature and searched in the experiments.

Weakly Interacting Massive particle~(WIMP) is an interesting candidate as its relic abundance can be determined by its interaction with SM particles in plasma in the early Universe~\cite{Schumann_2019}. A WIMP with the coupling of the order of electroweak interaction strength should have mass in the range of 10 GeV - 100 TeV. Although, WIMP has been searched in colliders, direct detection and indirect detection experiments, so far, it has not been found. Also, WIMP is heavily constrained by direct experimental searches~\cite{Lisanti:2016jxe,Schumann_2019}, therefore, to continue with WIMP, one has to find a way to evade the stringent direct detection bounds.

A pseudo scalar DM can naturally evade the strong direct detection constraints as it has derivative couplings which imply a momentum suppression in the tree-level DM-nucleon scattering matrix which vanishes in the non-relativistic limit~\cite{Gross_2017,Abe_2020,Abe_2021,Gola:2021abm,Okada_2021,Oda_2015,Das_2019,Das_2022}. Hence, it is interesting to seek a pseudo-scalar particle as a WIMP dark matter.  

In this article, we consider a generic $U(1)_X$ model for pseudo scalar DM and discuss its phenomenological implications in this study. The interesting aspect of the $U(1)_X$ models is that the three generations of right-handed neutrinos~(RHNs) are required to eliminate the gauge and mixed gauge-gravity anomalies \cite{Das:2016zue,Okada:2016tci,Bandyopadhyay:2017bgh,Das:2019pua}. The RHNs mix with active neutrinos of SM via type I seesaw mechanism~\cite{Minkowski:1977sc,Schechter:1980gr,Mohapatra:1979ia,Schechter:1981cv} to generate the required light neutrino masses and flavor mixing. A model which explains both the neutrino mass problem and the nature of dark matter would be a major step forward in High energy physics~\cite{Ma:2006km,Hirsch:2013ola,Merle:2016scw,Avila:2019hhv,Mandal:2021yph,Mandal:2019oth}.

The pseudo scalar is not protected by any symmetry therefore in order for it to be a DM candidate it must have a lifetime much greater than the age of the Universe. We study for the feasible parameter space allowed by lifetime constraint on our pseudo scalar particle then we scan for the allowed parameter space by relic and direct detection bounds while respecting several other theoretical and experimental constraints.


The paper is organized as follows. In sec.~\ref{sec:model}, we introduce the model and discuss the details of the new fields and their interactions. In sec.~\ref{sec:constraints} we discuss some of the relevant theoretical and experimental constraints. In sec.~\ref{sec:dark-matter} we discuss the relic density, direct detection and other phenomenologically relevant studies. Finally, in sec.~\ref{sec:conclusion}, we conclude the article.

\section{Model}
\label{sec:model}

We consider a BSM framework based on $U(1)_X$ gauge group which has been studied well in the literature~\cite{Okada:2016tci,Bandyopadhyay:2017bgh,Das:2019pua,Das:2022oyx}. The model has three RHNs~($N_R^i$) and two new scalars $\Phi,\chi$ additionally other than SM particles. The particle content and their respective charges are given in Table~\ref{tab1}, where the family index $i$ runs from 1 to 3. The $U(1)_X$ charges of the particles can be written in terms of only two charges, $x_H$ and $x_\Phi$, as shown in Table~\ref{tab1}. The detailed equations are derived from~\cite{Das:2016zue} and given in Appendix~\ref{anomaly}.
 
\begin{table}[t]
\begin{center}
\begin{tabular}{||c|ccc||c||}
\hline
\hline
            & SU(3)$_c$ & SU(2)$_L$ & U(1)$_Y$ & U(1)$_X$ \\[2pt]
\hline
\hline
&&&&\\[-12pt]
$q_L^i$    & {\bf 3}   & {\bf 2}& $\frac{1}{6}$ & 		 $\frac{1}{6}x_H + \frac{1}{3}x_\Phi$   \\[2pt] 
$u_R^i$    & {\bf 3} & {\bf 1}& $\frac{2}{3}$ &  	  $\frac{2}{3}x_H + \frac{1}{3}x_\Phi$ \\[2pt] 
$d_R^i$    & {\bf 3} & {\bf 1}& $-\frac{1}{3}$ & 	 $-\frac{1}{3}x_H + \frac{1}{3}x_\Phi$  \\[2pt] 
\hline
\hline
&&&&\\[-12pt]
$\ell_L^i$    & {\bf 1} & {\bf 2}& $-\frac{1}{2}$ & 	 $- \frac{1}{2}x_H - x_\Phi$  \\[2pt] 
$e_R^i$   & {\bf 1} & {\bf 1}& $-1$   &		$- x_H - x_\Phi$  \\[2pt] 
\hline
\hline
$N_R$   & {\bf 1} & {\bf 1}& $0$   &	 $- x_\Phi$ \\[2pt] 
\hline
\hline
&&&&\\[-12pt]
$H$         & {\bf 1} & {\bf 2}& $\frac{1}{2}$  &  	 $\frac{x_H}{2}$  \\ 
$\Phi$      & {\bf 1} & {\bf 1}& $0$  & 	 $2 x_\Phi$   \\ 
$\chi$      & {\bf 1} & {\bf 1}& $0$  & 	 $-x_\Phi$   \\ 
\hline
\hline
\end{tabular}
\end{center}
\caption{
Particle content of  the minimal U$(1)_X$ model where $i(=1, 2, 3)$ represents the family index.}
\label{tab1}
\end{table}
\subsection{Scalar Sector}
The scalar part of the Lagrangian is given by,
\begin{align}
 \mathcal{L}_{s}=(D^{\mu}H)^{\dagger}(D_{\mu}H)+(D^{\mu}\Phi)^{\dagger}(D_{\mu}\Phi)+(D^{\mu}\chi)^{\dagger}(D_{\mu}\chi)-V(H,\Phi,\chi),
\end{align}
here the covariant derivative can be defined as
\begin{align}
D_{\mu}=\partial_{\mu}-ig_{s}T^{a}G^{a}_{\mu}-igT^{a}W^{a}_{\mu}-ig_{1}Y B^1_{\mu}-ig_{X}Y_{X}B^2_{\mu},
\end{align}
The scalar potential $V(H,\Phi,\chi)$ is given by,
\begin{align}
 V(H,\Phi,\chi) = &\mu_{H}^{2}H^{\dagger}H+\mu_{\Phi}^{2}\Phi^{\dagger}\Phi+m_{\chi}^{2}\chi^{\dagger}\chi-\lambda_{H}(H^{\dagger}H)^{2}-\lambda_{\Phi}(\Phi^{\dagger}\Phi)^{2}
-\lambda_{\chi}(\chi^{\dagger}\chi)^{2}\nonumber \\
 &-\lambda_{H\Phi}(H^{\dagger}H)(\Phi^{\dagger}\Phi)\nonumber 
 -\lambda_{\Phi\chi}(\Phi^{\dagger}\Phi)(\chi^{\dagger}\chi)-\lambda_{H\chi}(H^{\dagger}H)(\chi^{\dagger}\chi)+(\kappa_{T}\Phi^{\dagger}\chi^2 + \text{h.c.})
\end{align}
We parameterize the scalar fields as,
\begin{align}
 H=\frac{1}{\sqrt{2}}
 \begin{pmatrix}
  -i(\phi_{1}-i\phi_{2}) \\
  v_H+R_1+i\phi_{3}  \\
 \end{pmatrix},\hspace{0.2cm}
 \chi=\frac{1}{\sqrt{2}}(v_{\chi}+R_2+i\eta_{\chi}),
 \hspace{0.2cm}
 \Phi=\frac{1}{\sqrt{2}}(v_{\Phi}+R_3+i\eta_{\Phi})
\end{align}
$v_H, v_{\Phi}$ and $v_{\chi}$ are the VEVs of Higgs doublet and SM singlets $\Phi$ and $\chi$ respectively. $w^{\pm}=\phi_{1}\mp i\phi_{2}$ would be the Goldstone boson of $W^{\pm}$, while $\phi_{3},\eta_{\Phi}$ and $\eta_{\chi}$ will mix to give $G^0, G'$ and $A$, that would be the Goldstone bosons of the $Z$ and $Z^{'}$ bosons, and the physical pseudo-Nambu Goldstone boson respectively.
The real scalars $R_1, R_2$ and $R_3$ are not the mass eigenstates due to mixing which implies following mass matrix 
\begin{align}
 M^2_R=
\begin{pmatrix}
 2\lambda_{H}v^2_H  &   \lambda_{H\chi}v_H v_{\chi}  &  \lambda_{H\Phi}v_H v_{\phi}  \\
 \lambda_{H\chi}v_H v_{\chi} &  2\lambda_{\chi}v^2_{\chi}  &    (-\sqrt{2}\kappa_T+\lambda_{\Phi\chi}v_{\phi})v_{\chi}  \\
 \lambda_{H\Phi}v_H v_{\phi} &  (-\sqrt{2}\kappa_T+\lambda_{\Phi\chi}v_{\phi})v_{\chi}  &  2\lambda_{\Phi}v^2_{\phi}+\frac{\kappa_T v^2_{\chi}}{\sqrt{2}v_{\phi}}  \\
\end{pmatrix}
\end{align}
The matrix $M^2_{R}$ can be diagonalized by an orthogonal matrix 
as follows~\cite{Darvishi_2021,Robens_2020}, 
\begin{align}
\mathcal{O}_{R} M_{R}^2 \mathcal{O}_R^T = 
\text{diag}(m_{h_1}^2,m_{h_2}^2,m_{h_3}^2),
\end{align}
 where
\begin{eqnarray}
 \label{rot1}
  \left( \begin{array}{c} h_1\\ h_2\\ h_3\\ \end{array} \right) = 
 \mathcal{O}_{R} \left( \begin{array}{c} R_1\\ R_2\\ R_3\\ 
 \end{array} \right). 
\label{eq:scalarmix}
 \end{eqnarray}
We assume the mass eigenstates to be ordered by their masses $m_{h_1}\leq m_{h_2}\leq m_{h_3}$. We will use the standard parameterization $\mathcal{O}_{R} = R_{23} R_{13} R_{12}$ where
\begin{equation}
R_{12} = \left(
\begin{array}{ccc}
c_{12} & s_{12} & 0\\
-s_{12} & c_{12} & 0\\
0 & 0 & 1
\end{array} \right), 
\quad R_{13} = \left(
\begin{array}{ccc}
c_{13} & 0 & s_{13}\\
0 & 1 & 0\\
-s_{13} & 0 & c_{13}
\end{array} \right), 
\quad R_{23} = \left(
\begin{array}{ccc}
1 & 0 & 0\\
0 & c_{23} &  s_{23}\\
0 & -s_{23} & c_{23}
\end{array} \right)
\end{equation}
$c_{ij} = \cos \theta_{ij}, s_{ij} = \sin \theta_{ij}$, where the angles $\theta_{ij}$ can be chosen to lie in the range $-\frac{\pi}{2}\leq\theta_{ij}\leq \frac{\pi}{2}$. The rotation matrix $\mathcal{O}_{R}$ is re-expressed in terms of the mixing angles in the following way: 
\begin{align}
\mathcal{O}_{R} = \left(
\begin{array}{ccc}
c_{12} c_{13}                          &  c_{13} s_{12}                        & s_{13}\\
-c_{23} s_{12} - c_{12} s_{13} s_{23}  & c_{23} c_{12} - s_{12} s_{13}  s_{23}  & c_{13} s_{23}\\
-c_{12} c_{23} s_{13} + s_{23} s_{12}  & -c_{23} s_{12} s_{13} - c_{12} s_{23} & c_{13} c_{23}
\end{array} \right).
\end{align}
It is possible to express the ten parameters of the scalar potential through the three physical Higgs masses, the three mixing angles, the three VEVs and $\kappa_T$~\cite{Robens_2020}. These relations are given by
\begin{equation}
    \begin{aligned}
        \lambda_H     & =\frac{1}{2 v_H^2} m_{h_i}^2 (\mathcal{O}_R^{i1})^2\,,        &
        \lambda_{\chi}        & =\frac{1}{2 v_\chi^2}  m_{h_i}^2 (\mathcal{O}_R^{i2})^2\,,      &
        \lambda_\Phi        & =\frac{1}{2 v_\Phi^2}  m_{h_i}^2 (\mathcal{O}_R^{i3})^2-\kappa_T\frac{v_\chi^2}{2\sqrt{2}v_\Phi^3}\,,        \\
        \lambda_{H \chi} & =\frac{1}{v_H v_\chi}  m_{h_i}^2 \mathcal{O}_R^{i1} \mathcal{O}_R^{i2}\,,   &
        \lambda_{H \Phi} & =\frac{1}{v_H v_\Phi}  m_{h_i}^2 \mathcal{O}_R^{i1} \mathcal{O}_R^{i3}\,,   &
        \lambda_{\Phi \chi}    & =\frac{1}{v_\Phi v_\chi}  m_{h_i}^2 \mathcal{O}_R^{i2} \mathcal{O}_R^{i3}+\sqrt{2}\frac{\kappa_T}{v_\Phi}\,,
    \end{aligned}
    \label{eq:quartic}
\end{equation}
The squared mass matrix for the CP-odd scalars in the weak basis ($\eta_{\chi}$, $\eta_{\Phi}$) is given as
\begin{align}
 M^2_{\eta}=\frac{\kappa_T}{\sqrt{2}}
\begin{pmatrix}
4v_{\Phi}  &    -2v_{\chi}  \\
-2v_{\chi}  & \frac{v^2_{\chi}}{v_{\Phi}} \\
\end{pmatrix}
\end{align}
This mass matrix can be diagonalized as $\mathcal{O}_\eta M^2_{\eta} \mathcal{O}_\eta^T=\text{diag}(0,m_A^2)$, where
\begin{align}
\mathcal{O}_\eta= 
\begin{pmatrix}
\cos\theta_{\eta} & \sin\theta_{\eta} \\
 -\sin\theta_{\eta} & \cos\theta_{\eta} \\
\end{pmatrix}, \,\, m^2_{A}=\frac{\kappa_T(4v^2_{\Phi}+v^2_{\chi})}{\sqrt{2}v_{\Phi}}
\end{align} 
and the null mass corresponds to the would-be Goldstone boson $G'$, which will be eaten by $Z'$.
The gauge eigenstates $(\eta_\chi\,\, \eta_\Phi)$ are related with the mass eigenstates $(G'\,\, A)$ as
\begin{align}
 \begin{pmatrix}
  \eta_{\chi} \\
  \eta_{\Phi}  \\ 
 \end{pmatrix}
=
\begin{pmatrix}
 \cos\theta_{\eta} & -\sin\theta_{\eta} \\
 \sin\theta_{\eta} & \cos\theta_{\eta} \\
\end{pmatrix}
\begin{pmatrix}
  G' \\
  A \\ 
\end{pmatrix},\text{  where  }  \tan\theta_{\eta} = \frac{2v_{\Phi}}{v_{\chi}}
\end{align}
Note that fixing one of the Higgs masses to the mass of the observed Higgs boson, $m_{h_1}=125$ GeV, and fixing the Higgs doublet vev to its SM value, $v_H= 246$ GeV,
leaves eight free input parameters from the scalar sector:
\begin{align}
m_{h_2},\,\, m_{h_3},\,\, m_A,\,\, \theta_{12},\,\, \theta_{13},\,\, \theta_{23},\,\, v_\chi \text{ and } v_{\Phi}
\end{align}

\subsection{Gauge Sector}

We note that a kinetic mixing can occur provided there are two or more field strength tensors $B^1_{\mu\nu}$ and $B^2_{\mu\nu}$ which are neutral under some gauge symmetry. This only arises for the abelian $U(1)$ gauge group. Thus in our case with the gauge group $SU(3)_c\otimes SU(2)_L\otimes U(1)_Y\otimes U(1)_X$, there can be kinetic mixing between two abelian gauge group $U(1)_Y$ and $U(1)_X$. The kinetic terms can be expressed as follows
\begin{align}
\mathcal{L}_{\rm Kinetic} = -\frac{1}{4}B^{1\mu\nu}B^1_{\mu\nu} - \frac{1}{4}B^{2\mu\nu}B^2_{\mu\nu} - \frac{\kappa}{2} B^{1\mu\nu}B^2_{\mu\nu},
\label{eq:kinetic-mixing}
\end{align}
where $B^1_{\mu\nu}$ and $B^2_{\mu\nu}$ are the filed strength tensors of the gauge groups $U(1)_Y$ and $U(1)_X$, respectively. The requirement of positive kinetic energy implies that kinetic coefficient $|\kappa|<1$. One can diagonalize the kinetic mixing term as follow
\begin{align}
 \begin{pmatrix}
  \tilde{B}_{1} \\
  \tilde{B}_{2}  \\ 
 \end{pmatrix}
=
\begin{pmatrix}
 1 & \kappa \\
 0 & \sqrt{1-\kappa^2} \\
\end{pmatrix}
\begin{pmatrix}
 B_{1} \\
 B_{2} \\
\end{pmatrix}
\end{align}
Let's first set kinetic mixing $\kappa=0$ to fix the notation. The diagonal component of the $SU(2)_L$ gauge filed $W_{3\mu}$ will mix with the $U(1)_Y$ and $U(1)_X$ gauge fields $B^1_{\mu}$ and $B^{2}_\mu$. To determine the gauge boson mass spectrum, we have to expand the scalar kinetic terms 
\begin{align}
\mathcal{L}_{s}=(D^{\mu}H)^{\dagger}(D_{\mu}H)+(D^{\mu}\Phi)^{\dagger}(D_{\mu}\Phi)+(D^{\mu}\chi)^{\dagger}(D_{\mu}\chi)
\end{align}
and have to replace the fields $H,\Phi$ and $\chi$ by the following expressions such as
\begin{align}
H=\frac{1}{\sqrt{2}}\begin{pmatrix}
0\\
v_H+R_1\end{pmatrix},\,\, \braket{\Phi}=\frac{1}{\sqrt{2}} (v_\Phi+R_2) \text{ and } \braket{\chi}=\frac{1}{\sqrt{2}} (v_\chi+R_3).
\end{align}
With this above replacement, we can expand the scalar kinetic terms $(D^\mu H)^\dagger (D_\mu H)$, $(D^\mu\Phi)^\dagger (D_\mu\Phi)$ and $(D^\mu\chi)^\dagger (D_\mu\chi)$ as follows
\begin{align}
& (D^{\mu}H)^{\dagger}(D_{\mu}H)\equiv\frac{1}{2}\partial^{\mu}R_1\partial_{\mu}R_1+\frac{1}{8}(R_1+v_H)^{2}\Big(g^{2}|W_{1}^{\mu}-iW_{2}^{\mu}|^{2}+(gW_{3}^{\mu}-g_{1}B^{1\mu}-\tilde{g}B^{2\mu})^{2}\Big)\\
& (D^{\mu}\Phi)^{\dagger}(D_{\mu}\Phi)\equiv\frac{1}{2}\partial^{\mu}R_2\partial_{\mu}R_2+\frac{1}{2}(R_2+v_\Phi)^{2}(2g_{1}^{'} B^{2\mu})^{2}\\
& (D^{\mu}\chi)^{\dagger}(D_{\mu}\chi)\equiv\frac{1}{2}\partial^{\mu}R_3\partial_{\mu}R_3+\frac{1}{2}(R_3+v_\chi)^{2}(g_{1}^{'} B^{2\mu})^{2}.
\end{align}
where we have defined $\tilde{g}=g_{X}x_H$ and $g_{1}^{'}=g_{X}x_\Phi$. SM charged gauge boson $W^{\pm}$ can be easily recognized with mass $M_{W}=\frac{g v_H}{2}$. On the other hand, the mass matrix of the neutral gauge bosons is given by
\begin{align}
\mathcal{L}_M=\frac{1}{2} V_0^T M_G^2 V_0
\end{align}
where
\begin{align}
V_0^T= \begin{pmatrix}B^1_\mu & W_{3\mu} & B^2_\mu \end{pmatrix} \text{   and    }
M_G^2=
\begin{pmatrix}
\frac{1}{4}g^2_1 v_H^2 &  -\frac{1}{4} g g_1 v_H^2   &   \frac{1}{4} g_1 \tilde{g} v_H^2 \\
-\frac{1}{4} g g_1 v_H^2  &   \frac{1}{4} g^2 v_H^2  &  -\frac{1}{4} g  \tilde{g} v_H^2  \\
\frac{1}{4} g_1  \tilde{g} v_H^2  &  -\frac{1}{4} g  \tilde{g} v_H^2  &  \frac{1}{4}  \tilde{g}^2 v_H^2+ 4  g_1^{'2} v_T^2  
\end{pmatrix}
\end{align}
Following linear combination of $B_1^{\mu}$, $W_{3}^{\mu}$ and $B_2^{\mu}$ gives definite mass eigenstates $A_0^{\mu}$, $Z_0^{\mu}$ and $Z_0^{'\mu}$~(when $\kappa=0$),
\begin{align}
 \begin{pmatrix}
  B_1^{\mu}\\
  W_{3}^{\mu}\\
  B_2^{\mu}
 \end{pmatrix}
= \begin{pmatrix}
 \cos\theta_{w}  & -\sin\theta_{w}~\cos\theta_0 &  \sin\theta_{w}~\sin\theta_0 \\
 \sin\theta_{w}  & \cos\theta_{w} \cos\theta_0  &  -\cos\theta_{w} ~\sin\theta_0 \\
 0                      & \sin\theta_0                        &  \cos\theta_0 \\
\end{pmatrix}
\begin{pmatrix}
 A_0^{\mu}\\
 Z_0^{\mu} \\
 Z_0^{'\mu} \\
\end{pmatrix}
\label{eq:gauge-tranformation}
\end{align}
where $\theta_{w}$ is the Weinberg mixing angle and,
\begin{align}
 \text{tan} 2\theta_0=\frac{2\tilde{g}\sqrt{g^{2}+g_{1}^{2}}}{\tilde{g}^{2}+16\left(\frac{v_T}{v_H}\right)^{2}g_{1}^{'2}-g^{2}-g_{1}^{2}} \text{ with } v_T=\frac{\sqrt{v_\chi^2+4 v_\Phi^2}}{2}.
\end{align}
When the kinetic mixing $\kappa$ is non-zero, the fields $A_0, Z_0$ and $Z'_0$ are not orthogonal. In the kinetic term diagonalized basis $\tilde{V}_0^T=(\tilde{B}_1\,\,W_3\,\,\tilde{B}_2)$, the mass matrix of the neutral gauge boson can be written as
\begin{align}
\mathcal{L}_M=\frac{1}{2} \tilde{V}_0^T S^T M_G^2  S \tilde{V}_0 = \frac{1}{2} \tilde{V}_0^T  \tilde{M}_G^2  \tilde{V}_0
\end{align}
where
\begin{align}
S=\begin{pmatrix} 
1  &  0  & -\frac{\kappa}{\sqrt{1-\kappa^2}}  \\
0  & 1   & 0 \\
0  &  0  & \frac{1}{\sqrt{1-\kappa^2}}
\end{pmatrix},
\tilde{M}_G^2=S^T  M_G^2  S = \begin{pmatrix}
\frac{1}{4}g_1 v_H^2 &  -\frac{1}{4} g g_1 v_H^2   &   \frac{1}{4} g_1 \tilde{g}_t v_H^2 \\
-\frac{1}{4} g g_1 v_H^2  &   \frac{1}{4} g^2 v_H^2  &  -\frac{1}{4} g  \tilde{g}_t v_H^2  \\
\frac{1}{4} g_1  \tilde{g}_t v_H^2  &  -\frac{1}{4} g  \tilde{g}_t v_H^2  &  \frac{1}{4}  \tilde{g}_t^2 v_H^2+ 4  g_1^{''2} v_T^2  
\end{pmatrix}
\end{align}
with $\tilde{g}_t=\frac{\tilde{g}-g_1\kappa}{\sqrt{1-\kappa^2}}$. Hence, comparing $M_G^2$ and $\tilde{M}_G^2$ we see that the overall effect of kinetic mixing introduction is just the modification of $\tilde{g}$ to $\tilde{g}_t$ and $g_1'$ to $g_1^{''}$. Hence, they can be related to orthogonal fields $A,Z$ and $Z'$ by the same transformation as in Eq.~\ref{eq:gauge-tranformation} but now $\theta_0$ is replaced by $\theta_{12}$
\begin{align}
 \begin{pmatrix}
  \tilde{B}_1^{\mu}\\
  W_{3}^{\mu}\\
  \tilde{B}_2^{\mu}
 \end{pmatrix}
= \begin{pmatrix}
 \cos\theta_{w}  & -\sin\theta_{w}~\cos\theta &  \sin\theta_{w}~\sin\theta \\
 \sin\theta_{w}  & \cos\theta_{w} \cos\theta  &  -\cos\theta_{w} ~\sin\theta \\
 0                      & \sin\theta                        &  \cos\theta \\
\end{pmatrix}
\begin{pmatrix}
 A^{\mu}\\
 Z^{\mu} \\
 Z^{'\mu} \\
\end{pmatrix}
\label{eq:final-gauge-tranformation}
\end{align}
where
\begin{align}
 \text{tan} 2\theta=\frac{2\tilde{g}_t\sqrt{g^{2}+g_{1}^{2}}}{\tilde{g}_t^{2}+16\left(\frac{v_T}{v_H}\right)^{2}g_{1}^{''2}-g^{2}-g_{1}^{2}} \text{  with  } \tilde{g}_t=\frac{\tilde{g}-g_1\kappa}{\sqrt{1-\kappa^2}} \text{ and } g_1^{''}=\frac{g_1'}{\sqrt{1-\kappa^2}}.
\end{align}
Masses of physical gauge bosons $A$, $Z$ and $Z^{'}$ are given by,
\begin{align}
 M_A=0,\,\,M_{Z,Z^{'}}^{2}=\frac{1}{8}\left(Cv_H^{2}\mp\sqrt{-D+v_H^{4}C^{2}}\right),
\end{align}
where,
\begin{align}
 C=g^{2}+g_{1}^{2}+\tilde{g}_t^{2}+16\left(\frac{v_T}{v_H}\right)^{2}g_{1}^{''2},
 \hspace{0.5cm}
 D=64v_H^{2}v_T^{2}(g^{2}+g_{1}^{2})g_{1}^{''2}.
\end{align}
The covariant derivative also can be expressed in terms of the orthogonal fields $\tilde{B}_1$ and $\tilde{B}_2$ as 
\begin{align}
D_{\mu}=\partial_{\mu}-ig_{s}T^{a}G^{a}_{\mu}-igT^{a}W^{a}_{\mu}-ig_{1}Y \tilde{B}^1_{\mu}-i\left(g_{X}Y_{X}\frac{1}{\sqrt{1-\kappa^2}}-g_1 Y \frac{\kappa}{\sqrt{1-\kappa^2}}\right)\tilde{B}^2_{\mu}
\end{align}
\subsection{Yukawa Sector}
The general form of the Yukawa interactions are given by, 
\begin{align}
 \mathcal{L}_{y} & =-y_{u}^{ij}\overline{q_{L}^{i}}\tilde{H}u^{j}_{R}-y_{d}^{ij}\overline{q_{L}^{i}}H d^{j}_{R}- y_{e}^{ij}\overline{\ell_{L}^{i}} H e^{j}_{R} - y_{\nu}^{ij}\overline{\ell_{L}^{i}}\tilde{H}\nu^{j}_{R}
 -\frac{1}{2} y_{M}^{ij}\Phi\overline{N_{R}^{ic}}N^{j}_{R} + \text{H.c.}
\label{Yukawa}
\end{align}
The last two terms are responsible for Dirac and Majorana masses of neutrinos. 
\subsection{Neutrino Mass}
Relevant light neutrino masses will come from the fourth and fifth terms of Eq.~\ref{Yukawa}. After the electroweak symmetry breaking, we can write the mass terms as,
 \begin{align}
  -\mathcal{L}_{M}=\sum_{i=1}^3\sum_{j=1}^3\overline{\nu_{jL}}m_{ij}^{D}N_{jR}+\frac{1}{2}\sum_{i,j=1}^3\overline{(N_{R})_{i}^{c}}M_{ij}^{R}N_{jR}+\text{H.c.},
 \end{align}
where $m_{ij}^{D}=\frac{y_{\nu}^{ij}v_H}{\sqrt{2}}$ and $M_{ij}^{R}=\frac{y_{M}^{ij}}{\sqrt{2}} v_{\Phi}$. Now we can write the $\mathcal{L}_{M}$ in the following matrix form,
\begin{align}
 -\mathcal{L}_{M}=\frac{1}{2}
 \begin{pmatrix}
  \overline{\nu_{L}} & \overline{(N_{R})^{c}} \\
 \end{pmatrix}
 \begin{pmatrix}
  {\bf 0}  &  m_{D} \\
  m_{D}^T  &  M_{R} \\
 \end{pmatrix}
\begin{pmatrix}
 (\nu_{L})^{c}\\
 N_{R} \\
\end{pmatrix} + \text{H.c.}
\label{eq:neutrino-mass}
\end{align}
In the seesaw approximation, this leads to the usual light neutrino mass matrix $m_{\nu}\approx -m_{D}M_R^{-1}m_D^T$. The $6\times 6$ matrix in Eq.~\ref{eq:neutrino-mass} is symmetric and can be diagonalized by the unitary matrix~(up to $\mathcal{O}(M_R^{-2})$)~\cite{Schechter:1981cv}
\begin{align}
\begin{pmatrix}
U & S \\
-S^{\dagger}  &  V
\end{pmatrix}^{\dagger}
\begin{pmatrix}
  {\bf 0}  &  m_{D} \\
  m_{D}^T  &  M_{R} \\
 \end{pmatrix}
 \begin{pmatrix}
U & S \\
-S^\dagger  &  V
\end{pmatrix}^*=
\begin{pmatrix}
\widehat{M_{\nu}} & {\bf 0} \\
{\bf 0}  &  \widehat{M_{N}}
\end{pmatrix}
\end{align}

where, $\widehat{M_{\nu}}=\text{diag}\{m_1,m_2,m_3\}$, $\widehat{M_N}=\text{diag}\{M_{N_1},M_{N_2},M_{N_3}\}$, $U\approx U_{\text{PMNS}}$, $V\approx {\bf I}$ and $S\approx m_D M_R^{-1}$.

\section{Theoretical and experimental constraints}
\label{sec:constraints}
We discuss the relevant theoretical and experimental constraints in this section.

\subsection{Vacuum Stability condition}
The scalar potential $V(H,\Phi,\chi)$ must be bounded from below~\cite{Kannike:2012pe} and it can be determined from the following symmetric matrix which comes from the quadratic part of the potential,
\begin{align}
 V^4_s=
\begin{pmatrix}
 \lambda_{H}    &   \frac{\lambda_{H\Phi}}{2}    &   \frac{\lambda_{H\chi}}{2}  \\
 \frac{\lambda_{H\Phi}}{2}  & \lambda_{\Phi}  &    \frac{\lambda_{\Phi\chi}}{2}  \\
 \frac{\lambda_{H\chi}}{2}  &  \frac{\lambda_{\Phi\chi}}{2}  & \lambda_{\chi}  \\
\end{pmatrix}
\end{align}
Above matrix will be positive-definite if following conditions are satisfied,
\begin{align}
 &\lambda_{H} > 0,
 \hspace{1cm}
 4\lambda_{H}\lambda_{\Phi}-\lambda_{H\Phi}^{2}>0 ,\nonumber \\
 &(-\lambda_{H}\lambda_{\Phi\chi}^{2}+\lambda_{H\Phi}\lambda_{\Phi\chi}\lambda_{H\chi}-\lambda_{\Phi}\lambda_{H\chi}^{2}+4\lambda_{H}\lambda_{\Phi}\lambda_{\chi}-\lambda_{H\Phi}^{2}\lambda_{\chi})>0.
\label{eq:stability}
\end{align}
An absolutely stable vacuum can be achieved if conditions given in Eq.~\eqref{eq:stability} are satisfied. The perturbativity condition on quartic coupling $\lambda_i\leq 4\pi$ and gauge coupling $g_X \leq \sqrt{4\pi}$ is being  ensured in the model.

\subsection{Invisible Higgs width constraint}
\label{sec:inv-Higgs}

In our model, $h_1$ scalar is SM Higgs by choice, which is the mixed state of three real scalar fields from eq. \ref{eq:scalarmix}. 
\begin{align}
h_{\text{SM}} = \sum_i \mathcal{O}_{R_{1i}} R_i
\label{eq:substitution}
\end{align}
If $m_{h_1} > m_{A}/2$, then the partial decay width to $AA$ is given by:
\begin{align}
\Gamma(h_1 \to A A) &= \frac{\lambda^2_{AAh_1}}{32\pi m_{h_1}}\sqrt{1-\frac{4m_{A}^2}{m_{h_1}^2}}
\end{align}
here $\lambda_{AAh_1}$ is trilinear coupling for vertex $A-A-h_1$. 
  
\begin{align*}
\lambda_{AAh_1} =  2 \sqrt{2} \kappa_T \mathcal{O}_{\eta_{21}}^\dagger \mathcal{O}_{\eta_{22}}^\dagger \mathcal{O}_{R_{12}} - & \left( \mathcal{O}_{\eta_{22}}^\dagger \right)^2 (\lambda_{H\Phi} v_H \mathcal{O}_{R_{11}} + \lambda_{\Phi\chi} v_{\chi} \mathcal{O}_{R_{12}} + 2 \lambda_{\Phi} v_{\phi} \mathcal{O}_{R_{13}}) \ - \\& \left( \mathcal{O}_{\eta_{21}}^\dagger \right)^2 (\lambda_{\Phi\chi} v_H \mathcal{O}_{R_{11}} + 2 \lambda_{\chi} v_{\chi} \mathcal{O}_{R_{12}} + (\sqrt{2} \kappa_T + \lambda_{\Phi\chi} v_{\phi}) \mathcal{O}_{R_{13}})
\end{align*}     
      
The invisible decay width of SM Higgs boson $h_1$ is given by $\Gamma^{\text{inv}}(h_1)=\Gamma(h_1\to A A)$. Then, one can calculate the invisible Higgs branching ratio as follow~\cite{Djouadi_2008,Robens_2020} 
\begin{align}
\text{BR}^{\text{inv}}(h_1)=
\frac{\Gamma^{\text{inv}}(h_1)}{\mathcal{O}^2_{R_{1i}}\Gamma^\text{SM}(h_1)+\Gamma^{\text{inv}}(h_1)},
\label{eq:inv-BR-Higgs}
\end{align}
here $\Gamma^\text{SM}(h_1)=4.1$ MeV. The current upper limit on the invisible Higgs branching ratio is from ATLAS experiment~\cite{ATLAS:2020kdi,ATLAS:2019cid,CMS:2018yfx},

\begin{align}
\text{BR}^{\text{inv}}(h_1) \leq 0.11.
\label{eq:invisible}
\end{align}

\subsection{Relic density constraint}
\label{sec:relic-constr}

The relic density bound is from Planck satellite data~\cite{Planck:2018vyg}.

\begin{align}
\Omega_{\text{DM}} h^2 = 0.12\pm 0.001
\label{eq:relic-density}
\end{align}

Any DM candidate must satisfy the relic bound given in equation~\ref{eq:relic-density}. 

\section{Dark Matter analysis}
\label{sec:dark-matter}

The general charge assignment for $U(1)_X$ model can be described in terms of only two free charges $x_H, ~x_{\phi}$. The charges $x_H$, $x_\Phi$ are the real parameters. For simplicity, one can fix $x_\Phi=1$ and vary $x_H$ only to characterize the $U(1)$ models. We can obtain $B-L$ case by choosing $x_H=0$. When $x_H=-2$, the left-handed fermions have no interactions with the $Z^\prime$ which leads to an $U(1)_R$ model. The interactions of $e_R$ and $d_R$ with $Z^\prime$ are absent when $x_H=-1$ and $1$ respectively. Thus, one can study any $U(1)$ models by changing only one parameter $x_H$ in our generic model. We have utilized this generality of our model in doing dark matter analysis. 

\vspace{0.5mm}
\textbf{Benchmark :} We have fixed the following independent parameters throughout the paper for the study of our pseudo scalar particle candidate to DM.

\begin{align*}
\textbf{BP :} \ \ \ \
x_{\phi}=1, \ \ \ \theta_{13,23}\approx 0, \ \ \ m_{h_3}=10^{13} ~ \text{GeV}, \ \ \ g_X=0.1, \ \ \ M_{Z'}=10^{14} ~ \text{GeV} 
\end{align*}
 
We have also set $m_{A}<<M_{N_i}$, therefore RHNs are not relevant for our DM analysis. Remaining free parameters such as $x_H$, $\theta_{12}$, $\kappa$, $m_{h_2}$, $v_{\phi}$, $v_{\chi}$ and $m_A$ are relevant for DM phenomenology.  

\begin{figure}[h!]
\includegraphics[height=3.5cm, width=5cm]{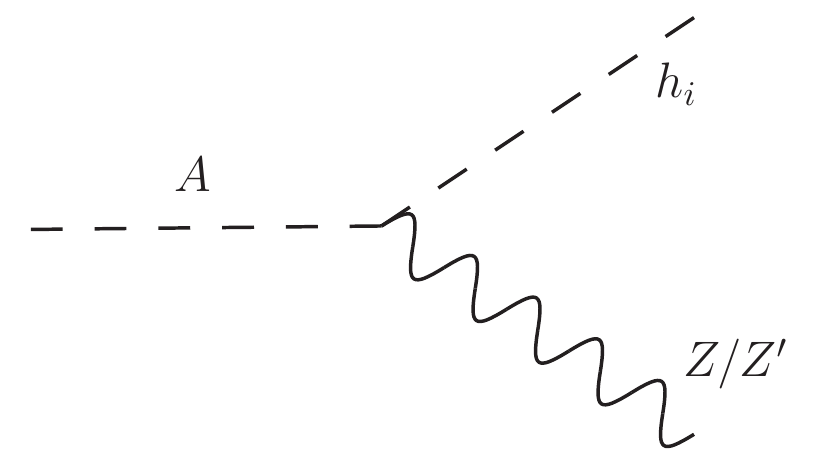} 
\hspace*{2cm}
\includegraphics[height=4cm, width=6cm]{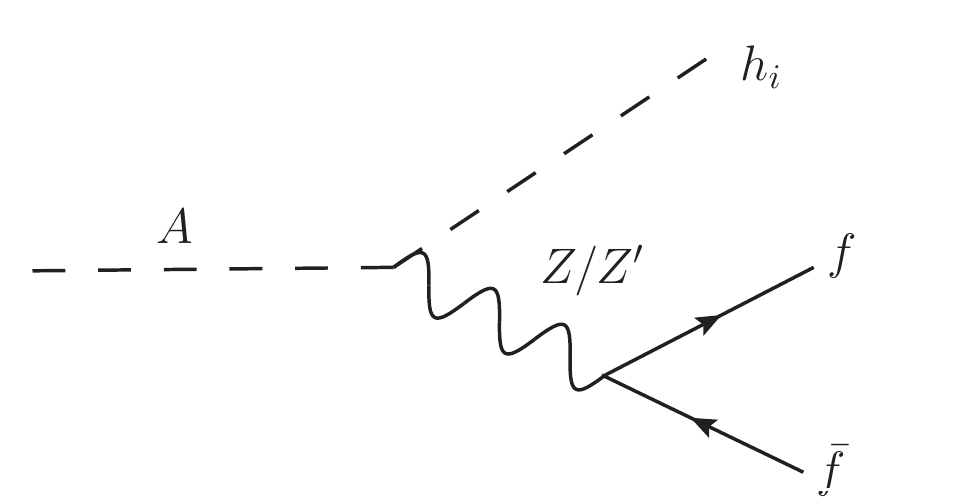}
\vspace*{1cm}
\includegraphics[height=4cm, width=6cm]{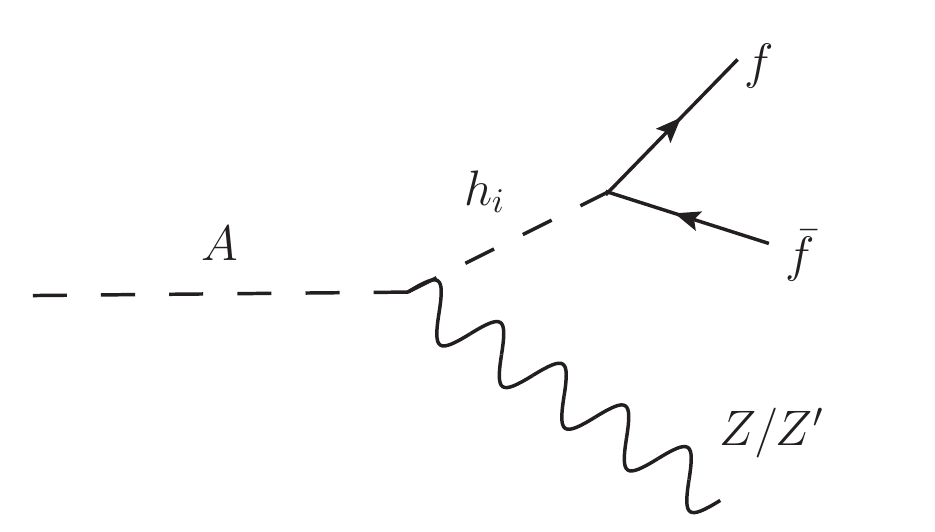}
\caption{The relevant two and three body decay modes for pseudo scalar DM are shown here.} 
\label{fdiagrams}
\end{figure}

\subsection{DM lifetime study}
\label{lifetime}

A good candidate for DM must be stable or have a lifetime greater than the age of the Universe. A conservative limit on the lifetime of dark matter is $\tau_{DM} \geq 10^{27}$ sec or in terms of decay width $\Gamma_{DM} \leq 6.6 \times 10^{-52}$ GeV, from the gamma rays observation of dwarf spheroidal galaxies~\cite{Baring_2016}. Our pseudo scalar should also follow this criterion in order to be a good DM candidate.  

There are two possible, two body decay modes namely $A \rightarrow \nu\nu, ~h_i Z$. Particles $Z'$, $N_i$ are chosen to be very heavy such that $h_i Z'$, $N N$ decay modes are not relevant to our study of DM phenomenology. The $\nu\nu$ channel is strongly suppressed due to the smallness of neutrino mass. The only two body decay channel is $h_i Z$, however, $h_3$ is also a massive particle. Hence, only $h_1 Z, ~h_2 Z$ modes are kinematically allowed in our interesting mass range of DM.

Apart from the two body decays, there could be two possible, three body decays such as $A \rightarrow Z f \bar{f}, \ h_i f \bar{f}$ are also relevant and could be dominant or subdominant depends on the kinematics. The former process contributes to non-zero gauge kinetic mixing. The latter process is mediated by $Z, ~Z'$ and a dominant three-body decay when kinematically allowed. We have shown the Feynman diagrams for these decay modes in figure~\ref{fdiagrams} and calculated the decay width expressions for all these modes in the following equations

\begin{figure}[h!]
\includegraphics[height=5.5cm, width=8cm]{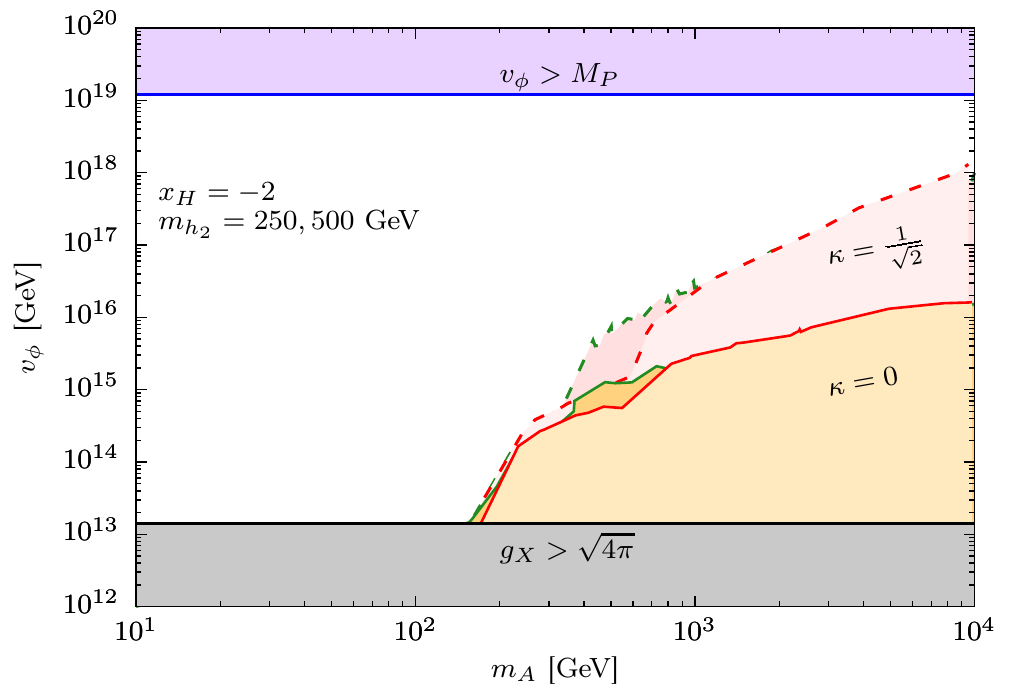}
\includegraphics[height=5.5cm, width=8cm]{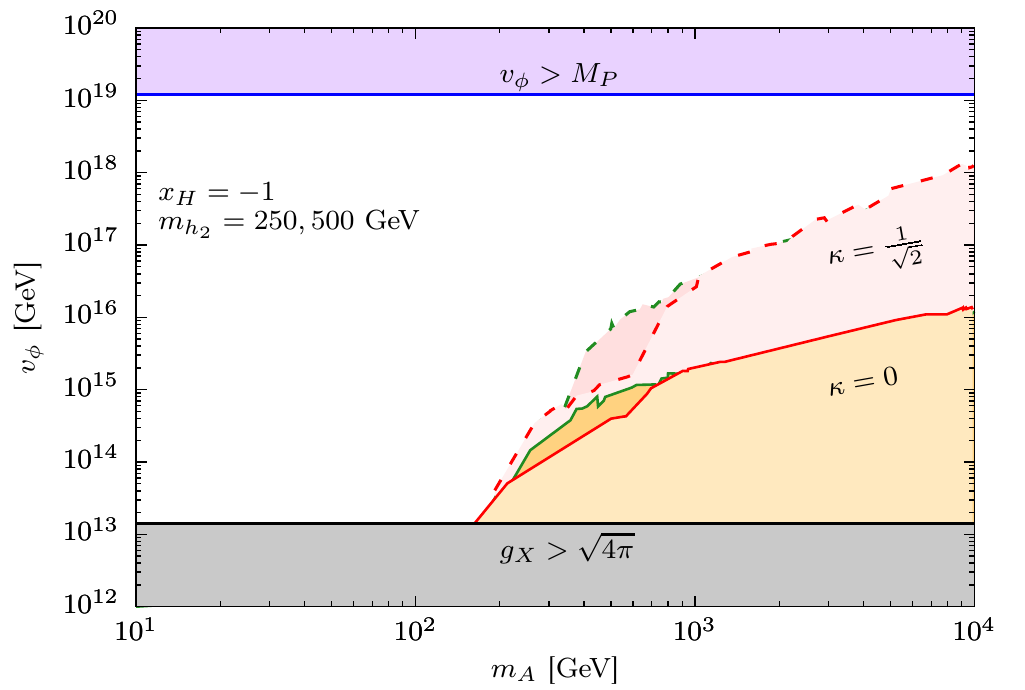}
\includegraphics[height=5.5cm, width=8cm]{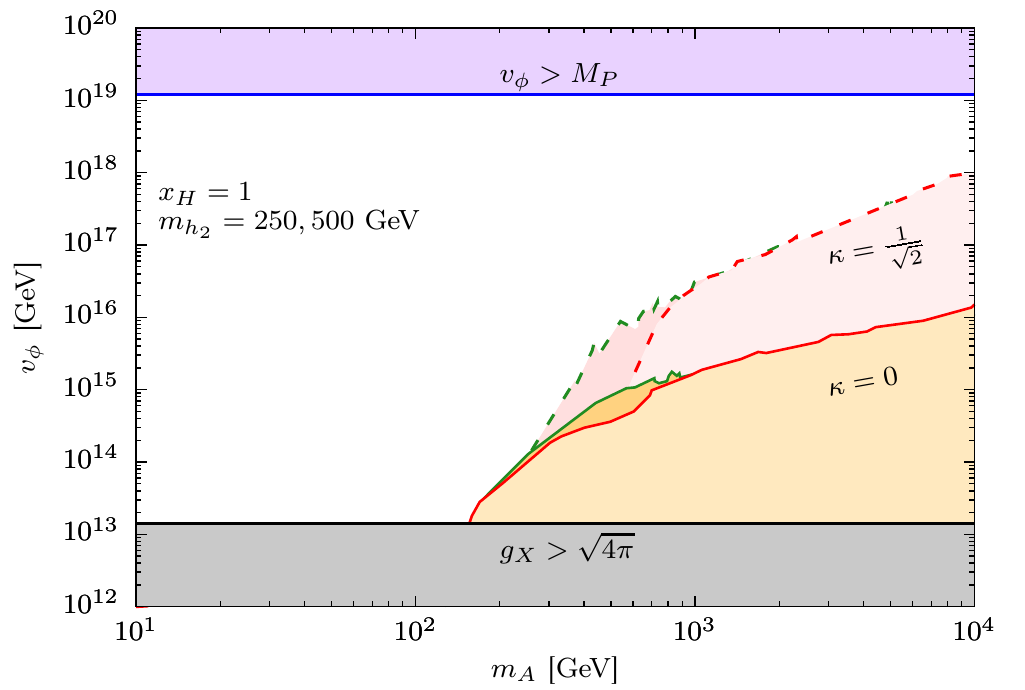}
\includegraphics[height=5.5cm, width=8cm]{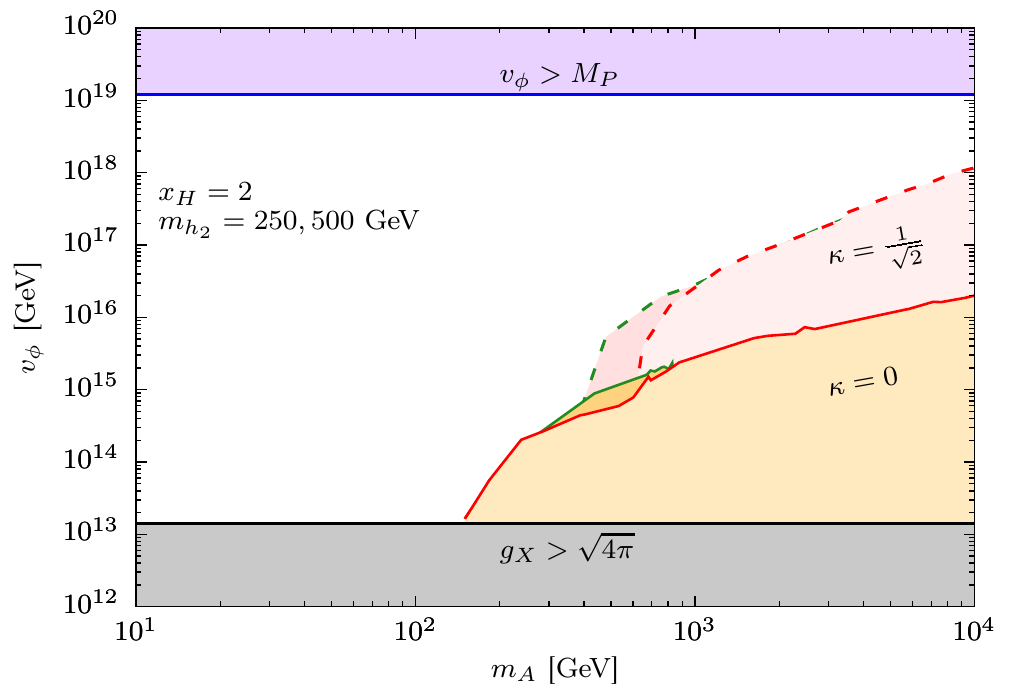}
\vspace{2mm}
\caption{Plots are displayed for $x_H = -2,-1,1,2$. In each plot $\theta_{12}=0.1$ is fixed. Allowed parameter space for a pseudo scalar is shown in $(m_A, v_{\phi})$ plane. The coloured region is disfavored. The purple, pink, orange and grey region is not allowed by Planck mass limit on vev, DM lifetime constraint~($\tau > 10^{27}$ sec)~\cite{Baring_2016} and perturbative unitarity bound on gauge coupling $g_X$. The orange~(pink) region corresponds to kinetic mixing 0~($1/\sqrt{2}$). The dark~(light) orange and pink region is correspond to $m_{h_2}=250~(500)$ GeV.} 
\label{width1}
\end{figure}

\begin{align}
& \Gamma(A\to h_i X)=\frac{|\lambda_{AXh_i}|^2 m_A^3}{16\pi M_{X}^2}\lambda^{\frac{3}{2}}\Big(1,\frac{M_X^2}{m_A^2},\frac{m_{h_i}^2}{m_A^2}\Big),
& \Gamma(A\to N_i N_i)=\frac{\cos^2\theta_\eta}{8\pi v_\Phi^2}m_A M_{N_i}^2 \sqrt{1-4\frac{M_{N_i}^2}{m_A^2}}
\label{dwidth1}
\end{align}
\begin{align}
& d\Gamma(A\to h_i f\bar{f})=\frac{m_A^5}{384 \pi^3}\Big[\big(|A_L^{if}(z)|^2+|A_R^{if}(z)|^2\big) \big\{\lambda\big(1,x_{h_i}^2,z\big)+\frac{x_f^2}{z}\big(2(1-x_{h_i}^2)^2+2(1+x_{h_i}^2)z-z^2\big)\big\}\nonumber \\
& -6\text{Re}\big(A_L^{if*}(z) A_R^{if}(z)\big) x_f^2 (2(1+x_{h_i}^2)-z) \Big]\frac{dz}{z}\lambda^{\frac{1}{2}}\big(1,x_{h_i}^2,z\big) \lambda^{\frac{1}{2}}\big(1,x_{f}^2,x_f^2\big);\,\,\, 4 x_f^2 \leq z \leq (1-x_{h_i})^2,
\label{dwidth2}
\end{align}
\begin{align}
\Gamma(A\to X f\bar{f})=\frac{|H_X(z)|^2 m^9 x_f^2}{128\pi^3 M_X^2 v_H^2} \big(z-4x_f^2\big)\lambda^{\frac{3}{2}}\big(1,\frac{M_X^2}{m_A^2},z\big)\frac{dz}{z} \lambda^{\frac{1}{2}}\big(1,x_{f}^2,x_f^2\big)
\label{dwidth3}
\end{align}
here $X=Z,~Z'$ and definitions of all other symbols used are given in appendix~\ref{dwidths}. The relevant parameters for the study of DM lifetime constraints on our pseudo scalar DM are U(1)$_X$ charge $x_H$, scalar mixing $\theta_{12}$ between Higgs h$_1$ and h$_2$, vev $v_{\phi}$ and $v_{\chi}$, gauge kinetic mixing $\kappa$ and masses $m_A$, $m_{h_2}$. Other free parameters are the same as in the benchmark. 

\begin{figure}[h!]
\includegraphics[height=5.5cm, width=8cm]{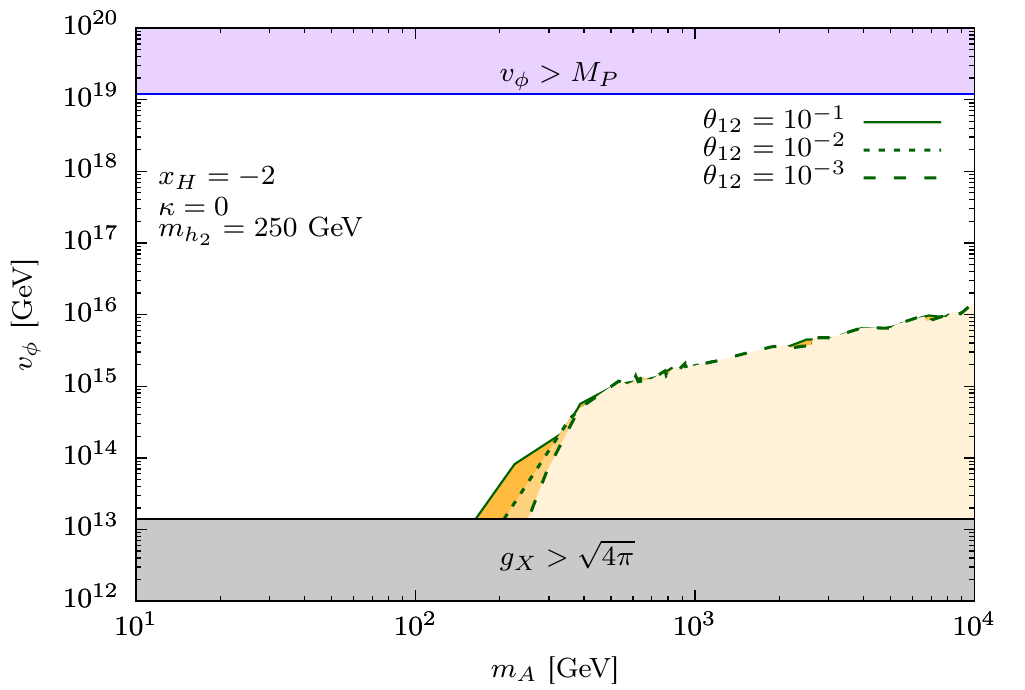}
\includegraphics[height=5.5cm, width=8cm]{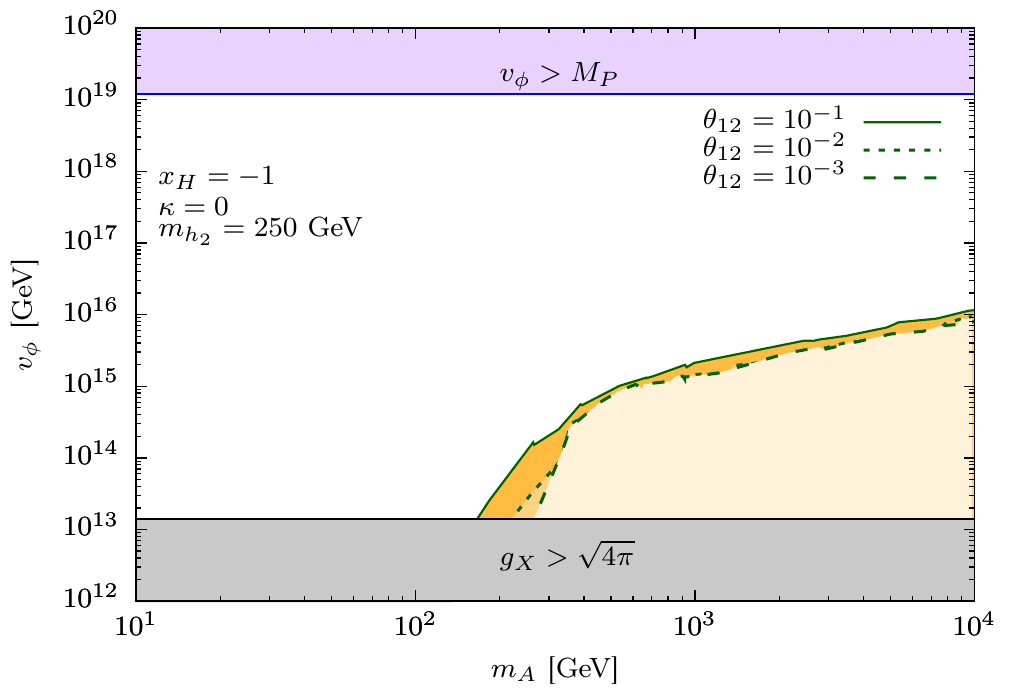}
\includegraphics[height=5.5cm, width=8cm]{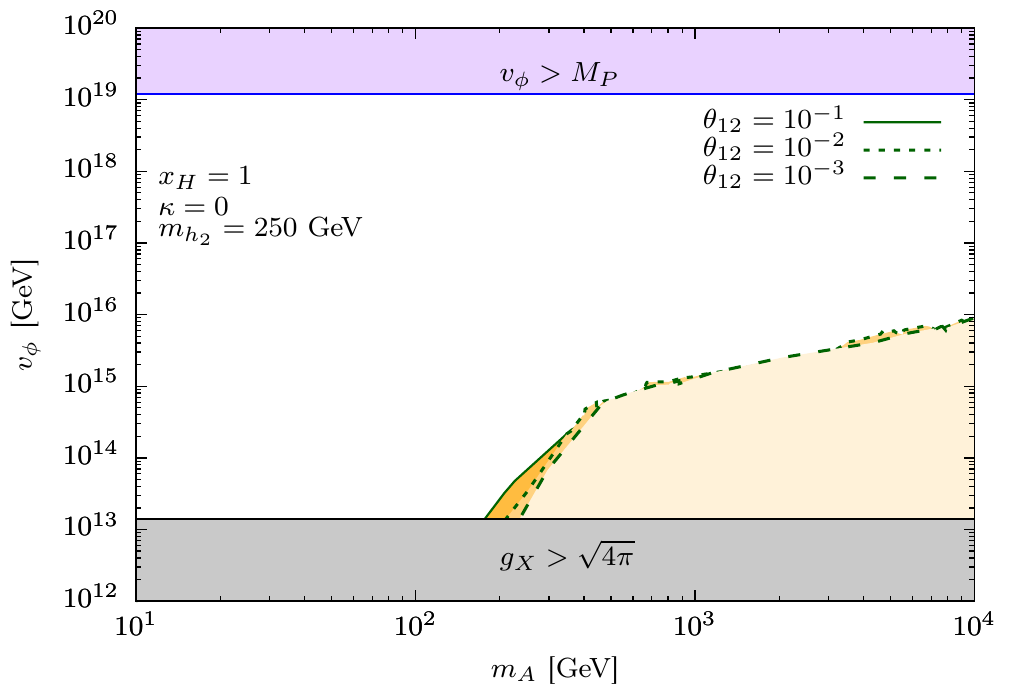}
\includegraphics[height=5.5cm, width=8cm]{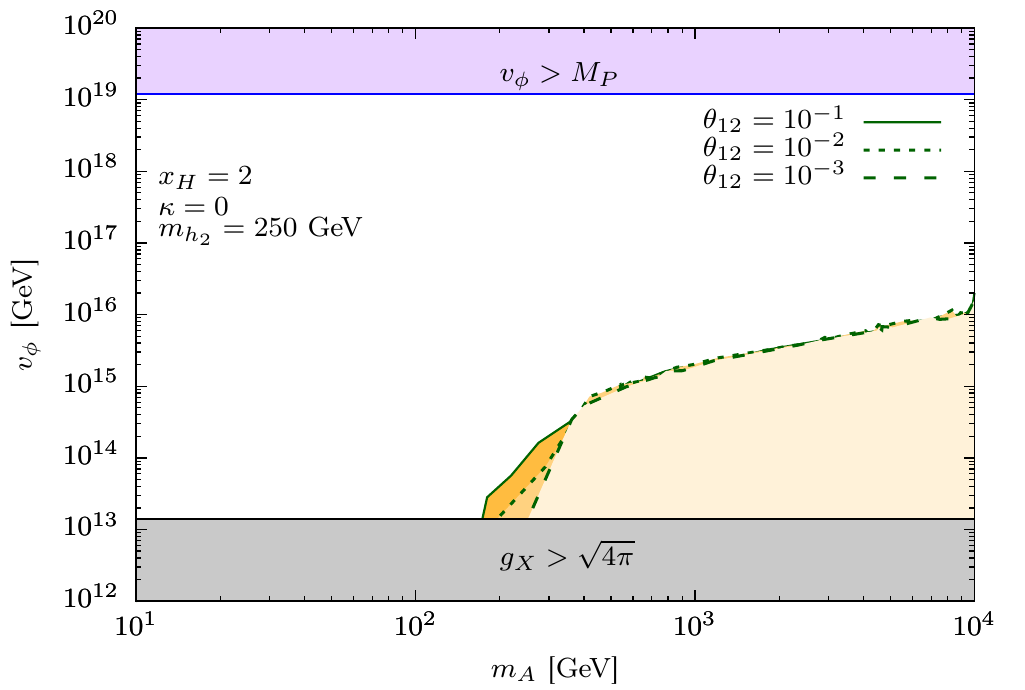}
\vspace{2mm}
\caption{Plots are displayed for $x_H = -2,-1,1,2$. Allowed parameter space for a pseudo scalar is shown in $(m_A, v_{\phi})$ plane. The disfavored region is coloured. Purple, orange and gray region is disfavored by Planck mass limit on vev, lifetime bound~\cite{Baring_2016} and perturbative unitarity limit on quartic coupling $\lambda_{\chi}$. The orange region~(from darker to lighter) is correspond to mixing $\theta_{12}=10^{-1,-2,-3}$.} 
\label{width2}
\end{figure}

\begin{figure}[h!]
\includegraphics[height=5.5cm, width=8cm]{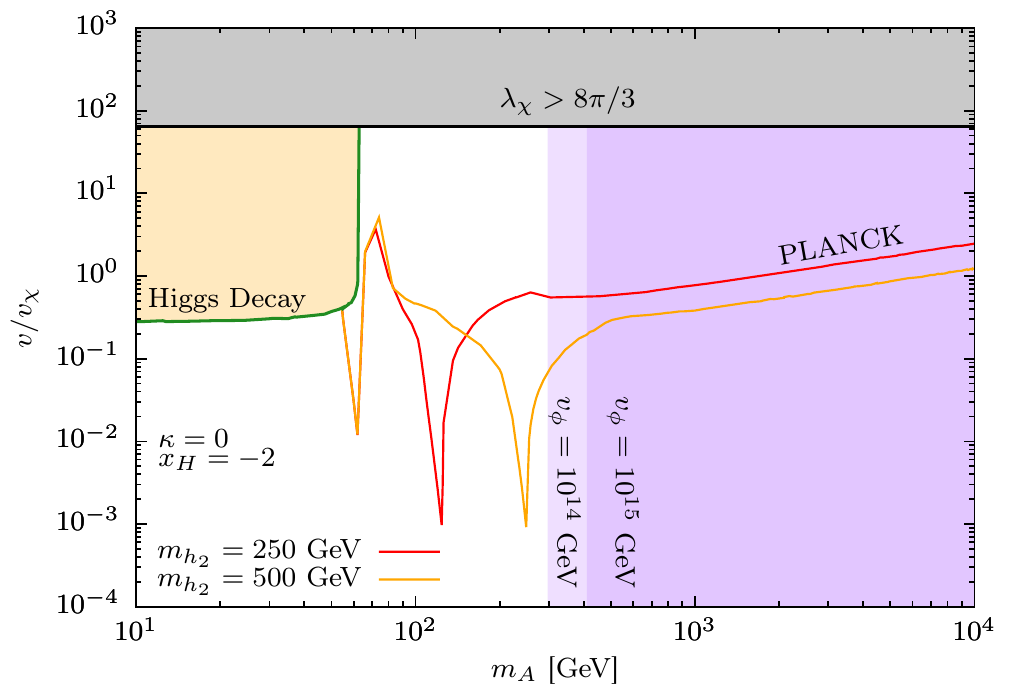}
\includegraphics[height=5.5cm, width=8cm]{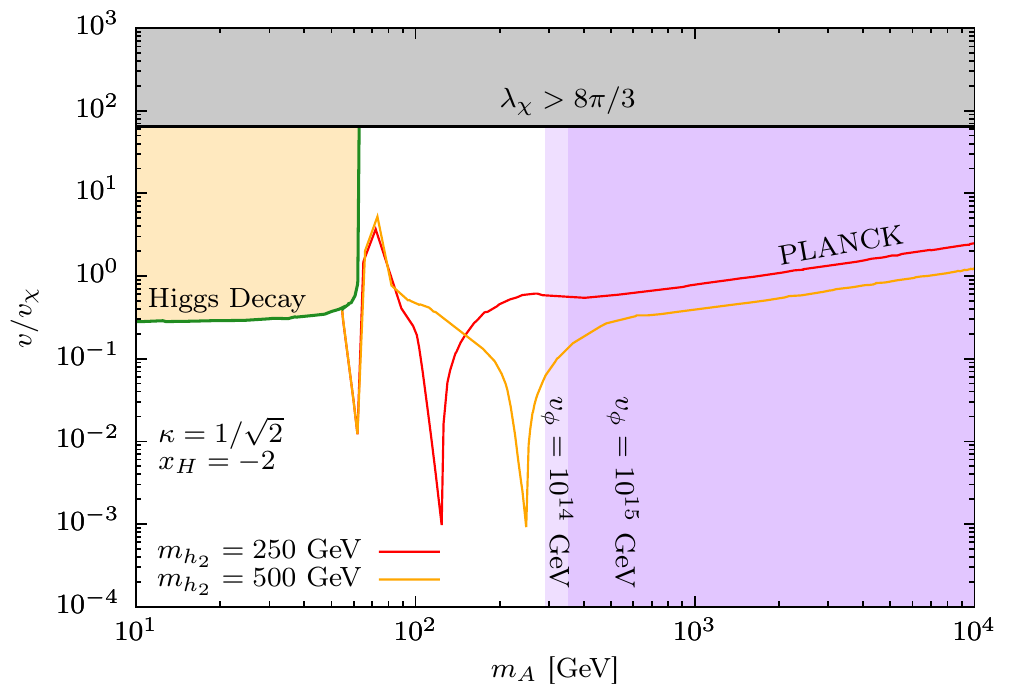}
\includegraphics[height=5.5cm, width=8cm]{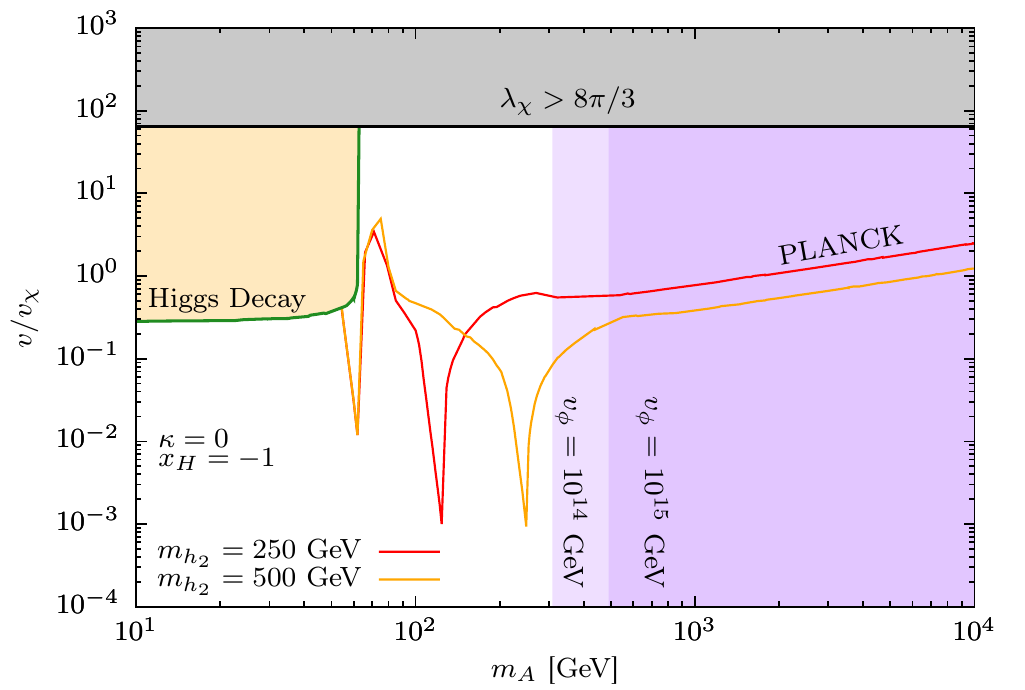}
\includegraphics[height=5.5cm, width=8cm]{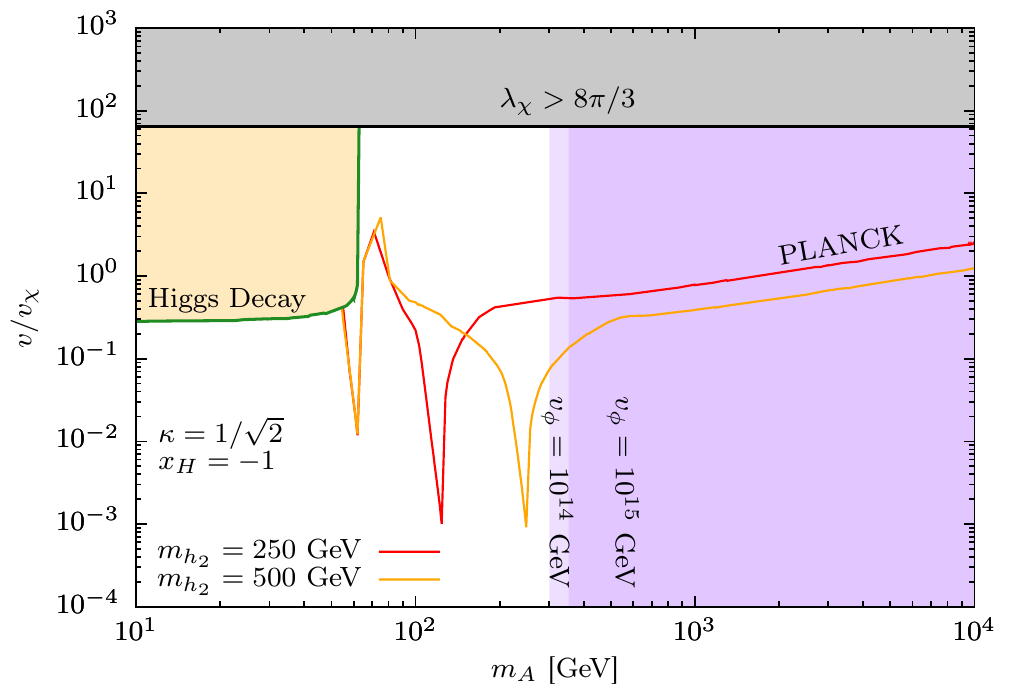}
\includegraphics[height=5.5cm, width=8cm]{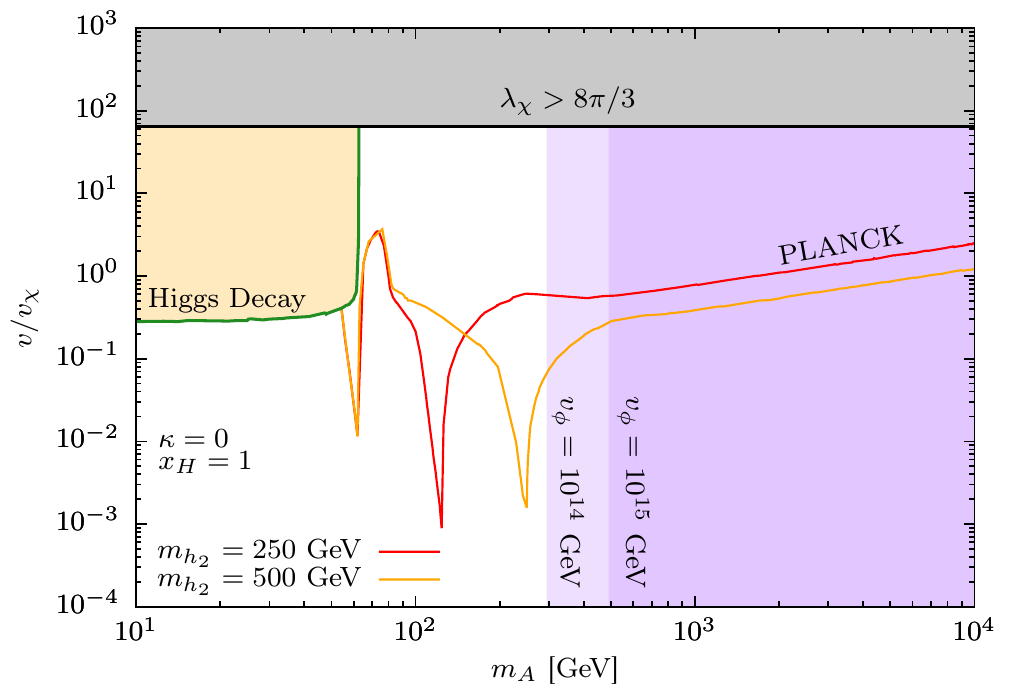}
\includegraphics[height=5.5cm, width=8cm]{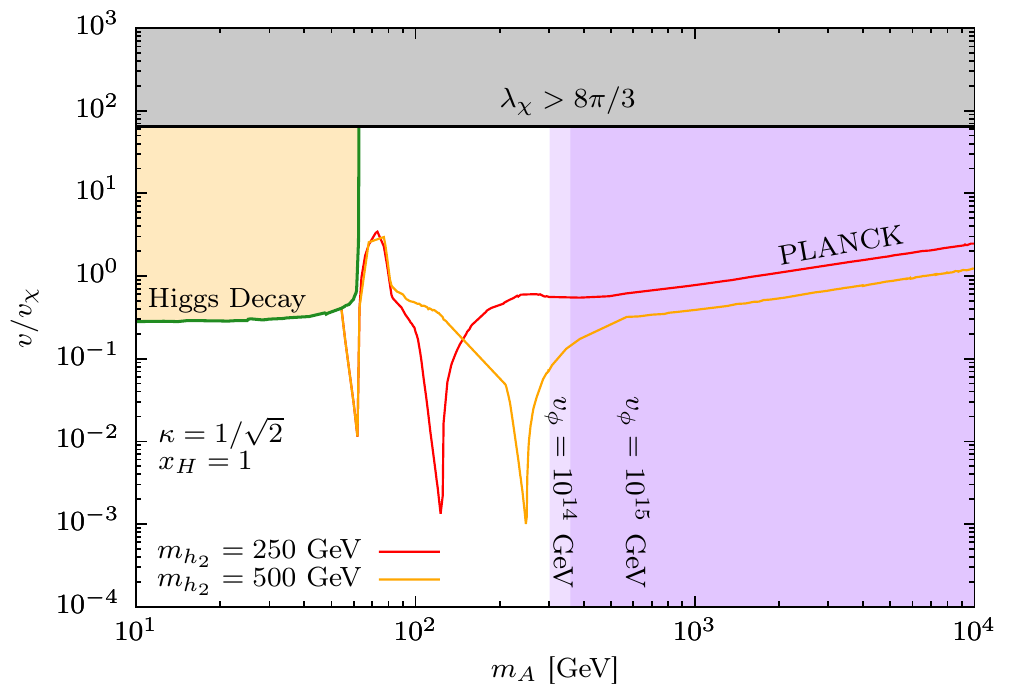}
\includegraphics[height=5.5cm, width=8cm]{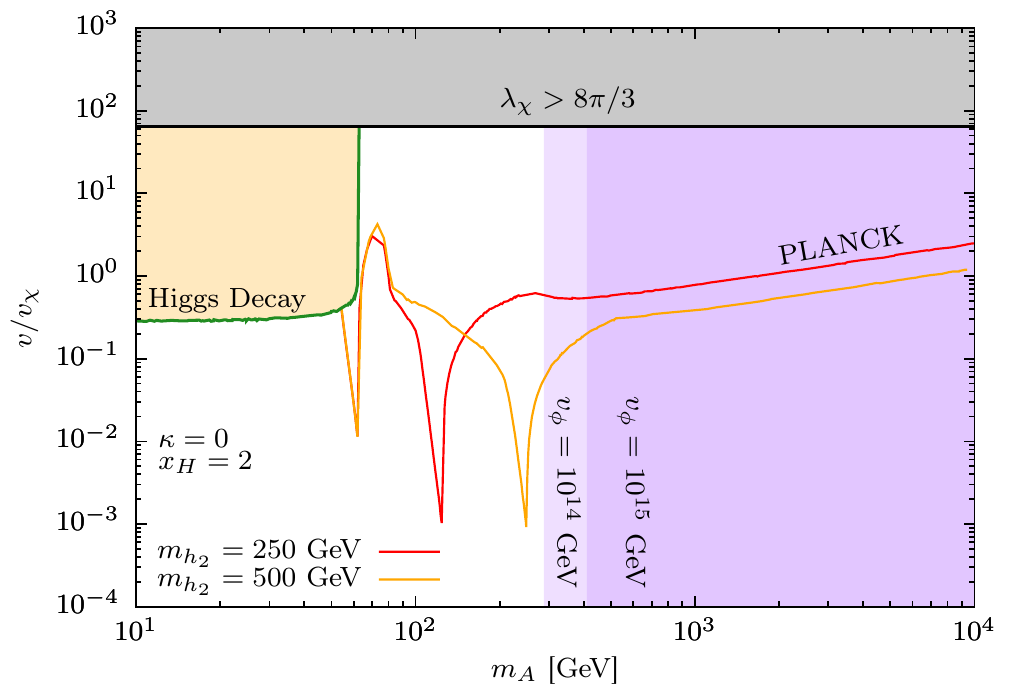}
\includegraphics[height=5.5cm, width=8cm]{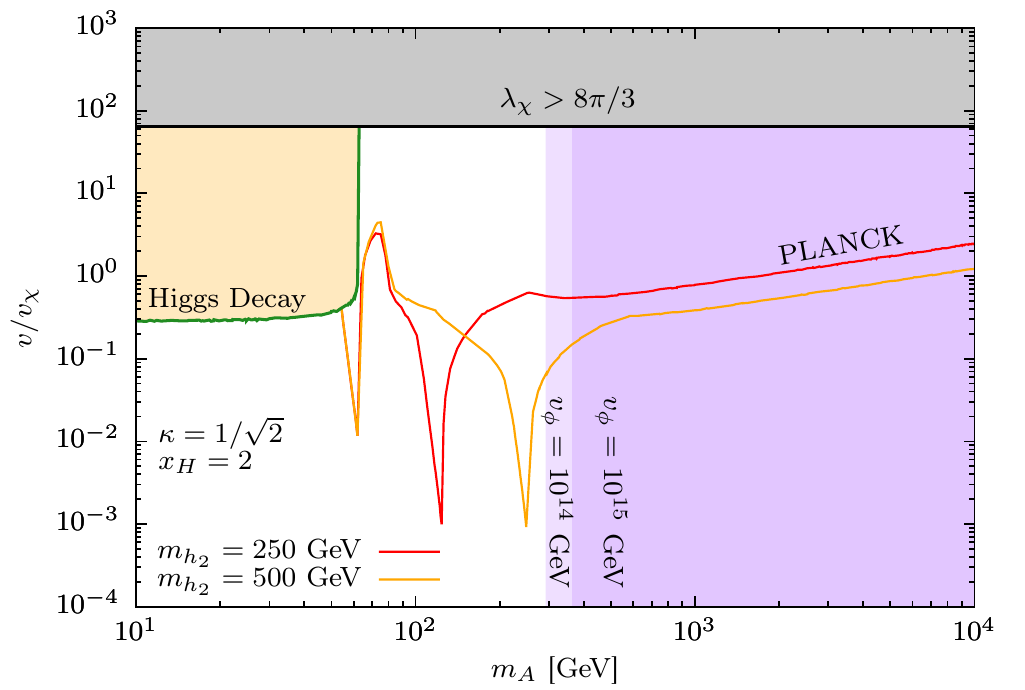}
\vspace{2mm}
\caption{Plots are displayed for $x_H = -2,-1,1,2$. Plots in left~(right) column are for $\kappa=0~(1/\sqrt{2})$. Allowed parameter space are shown in $(m_A, v/v_{\chi})$ plane. The disfavored region is coloured. The grey, orange and purple region is disfavored by perturbative unitarity bound on quartic coupling~$\lambda_{\chi}$, invisible Higgs width constraint~\ref{eq:invisible} and the conservative lifetime bound~($\tau > 10^{27}$ sec)~\cite{Baring_2016} for $v_{\phi}=10^{14,15} $ GeV. Red and orange curves represent the correct thermal relic abundance satisfied given by Planck data~\ref{eq:relic-density}. The two dips in red~(orange) curve at 62.5, 125~(250) GeV are the resonances due to $h_1$ and $h_2$.} 
\label{relic2}
\end{figure}

In fig~\ref{width1}, we show the allowed parameter space for DM in $(m_A, v_{\phi})$ plane. The four plots corresponding to $x_H=-2,-1,1,2$ from top left to right bottom. We have fixed mixing angle $\theta_{12}=0.1$ in all the plots here. The purple region on top of each plot is not allowed as vev $v_{\phi}$ gets larger than the Planck scale $M_P=1.2 \times 10^{19}$ GeV. The grey region on the bottom is disfavored by the perturbative unitarity bound of coupling $g_X > \sqrt{4\pi}$. The pink and orange region in the middle of each plot corresponds to kinetic mixing $\kappa=0,~1/\sqrt{2}$ respectively and it is disfavored by lifetime constraint. The light~(dark) orange and red region is due to $m_{h_2}=250~(500)$ GeV. The decay modes $h_i f \bar{f}, ~h_i Z$ are active in both pink and orange regions when allowed kinematically. However, in the pink region, where, we turn on the kinetic mixing, which opens up an additional decay mode $Z f \bar{f}$, implies the lesser parameter space available by lifetime bound on $(m_A, v_{\phi})$ plane. One can also see vev, $v_{\phi} > 10^{13}$ GeV is favored for DM mass around $\sim$ 100 GeV.

In fig~\ref{width2}, we have done the analysis similar to above, however, we have fixed $m_{h_2}=250$ GeV, gauge kinetic mixing $\kappa=0$ and vary the scalar mixing $\theta_{12}$. The U(1)$_X$ charge $x_H$ is labelled in each plot. The orange region~(from darker to lighter) is correspond to mixing $\theta_{12}=10^{-1,-2,-3}$. Here, channels $h_i f \bar{f}$ and $h_i Z$ are the only relevant decay modes. The disallowed region increases as one increase the scalar mixing due to decay modes $h_1 Z, ~h_2 Z$ both contributing significantly. The difference is significant only in a very small region of parameter space simply because of kinematics.   
 
\subsection{Relic density analysis}
\label{relic}

We can now study the relic density constraint on our pseudo scalar candidate to DM, which, we did by micrOMEGAs package~\cite{Belanger:2018ccd}. In fig~\ref{relic2}, we show the relic density analysis for our pseudo scalar DM in $(m_A, ~v/v_{\chi})$ plane. Plots in the left and right column correspond to gauge kinetic mixing $\kappa=0,~1/\sqrt{2}$ respectively. The scalar mixing $\theta_{12}=0.1$ is fixed in each plot. The U(1)$_X$ charge $x_H$ is also labelled in each plot. The grey region on top of each plot is disfavored by perturbative unitarity of quartic coupling $\lambda_{\chi} < 8\pi/3$. The light and dark purple region is disfavored by lifetime bound and it corresponds to $v_{\phi}=10^{14,15}$ GeV respectively. The light orange region in the middle is not allowed by invisible Higgs width constraint~\ref{eq:invisible}. The red and orange curves correspond to $m_{h_2}=250,~500$ GeV respectively and it represents the correct relic abundance following the Planck data~\ref{eq:relic-density}. The main annihilation channels that contribute to relic abundances are $ A~A \longrightarrow W^+W^-, ~ZZ, ~h_1h_1, ~h_2h_2 $. The two dips in each plot are the two resonances due to two Higgs poles at $m_{h_1}/2$ and $m_{h_2}/2$. One can see the allowed parameter space in $(m_A, ~v/v_{\chi})$ plane can be increased by increasing vev $v_{\phi}$ and decreasing gauge kinetic mixing $\kappa$. 
      
\subsection{Direct detection}
\label{direct}

\begin{figure}[h!]
\includegraphics[height=4.5cm, width=5.8cm]{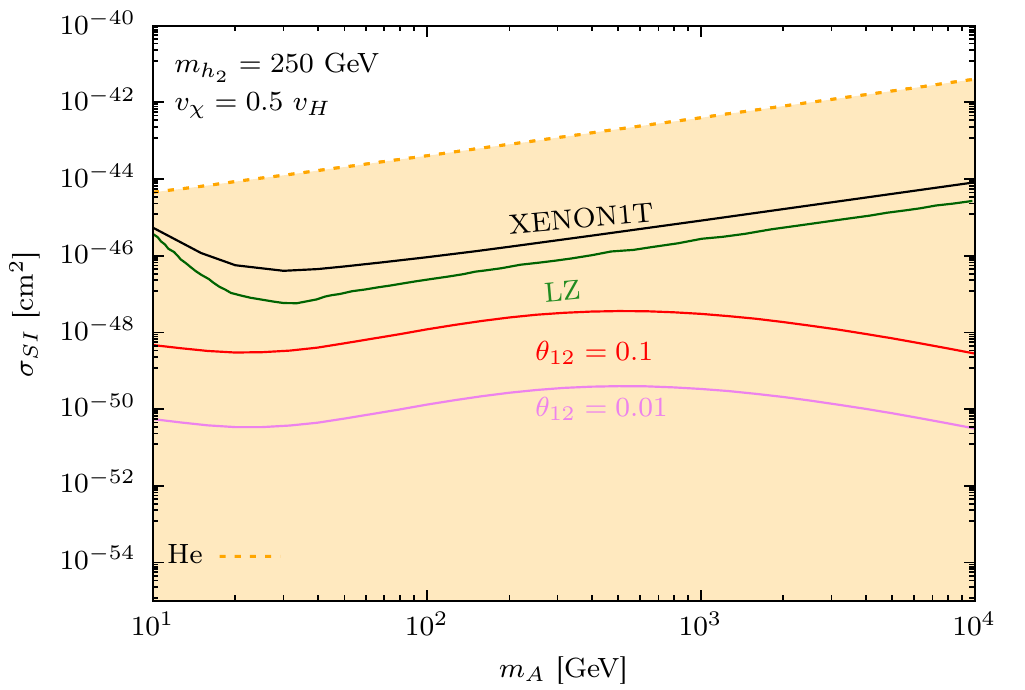}
\includegraphics[height=4.5cm, width=5.8cm]{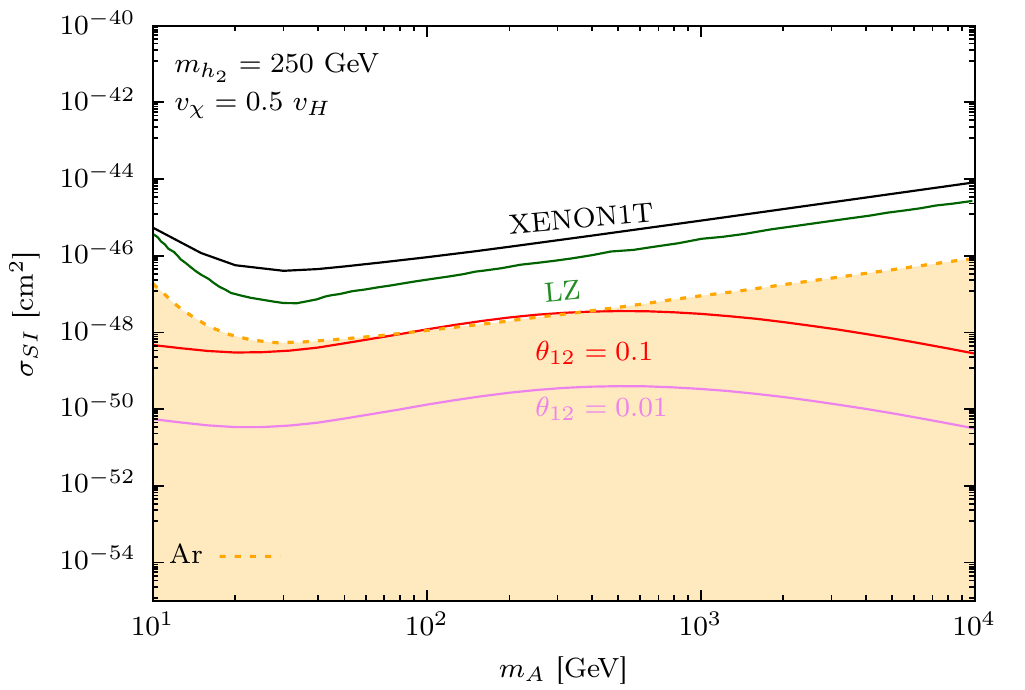}
\includegraphics[height=4.5cm, width=5.8cm]{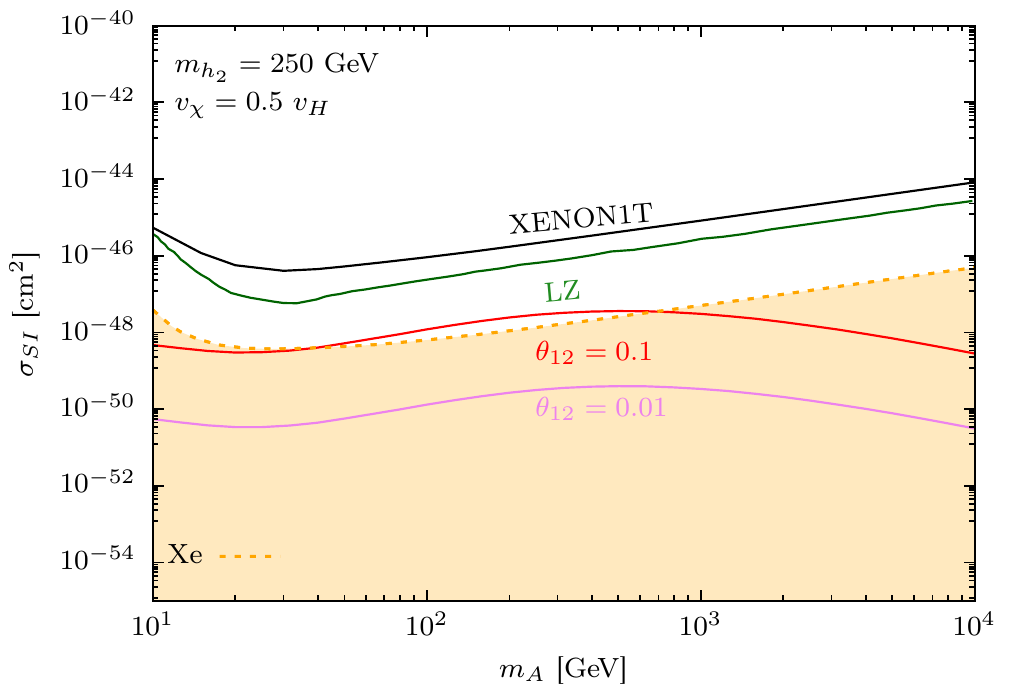}
\includegraphics[height=4.5cm, width=5.8cm]{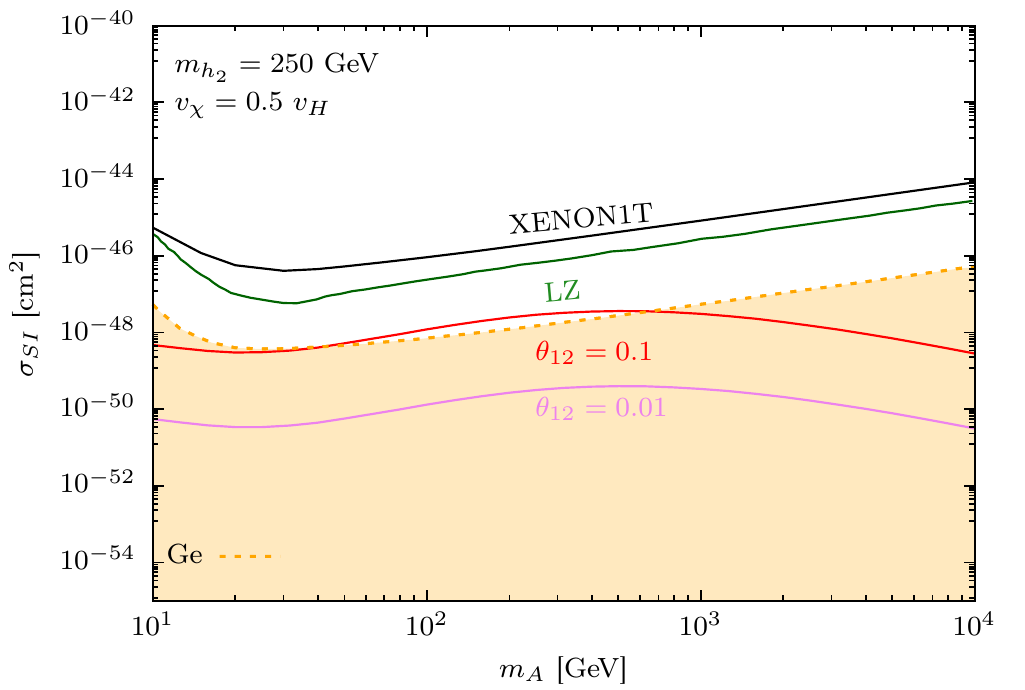}
\includegraphics[height=4.5cm, width=5.8cm]{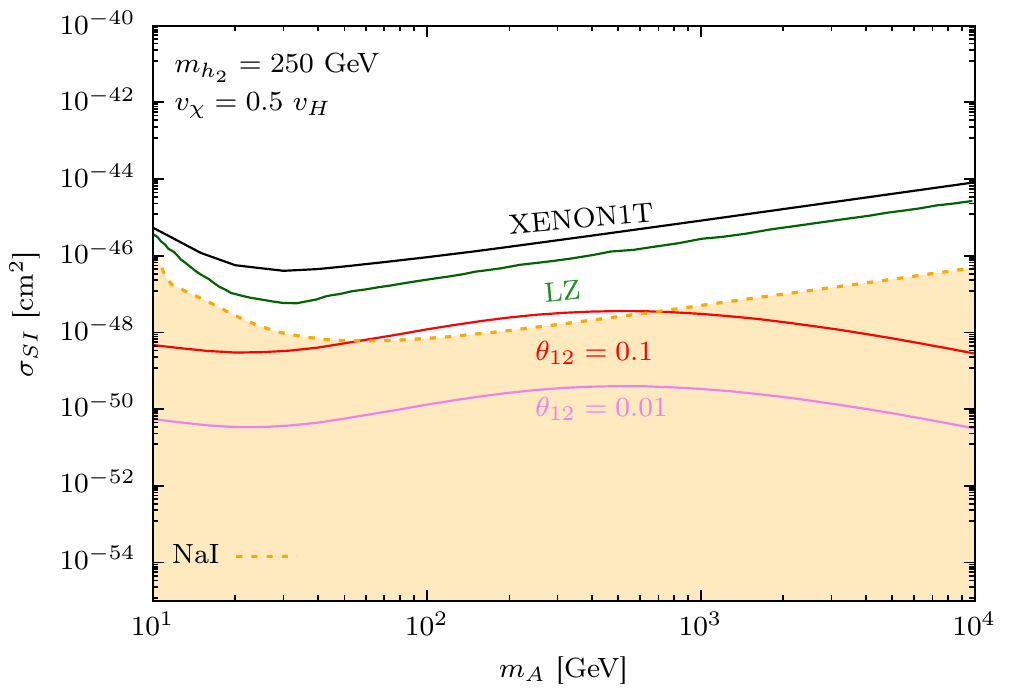}
\includegraphics[height=4.5cm, width=5.8cm]{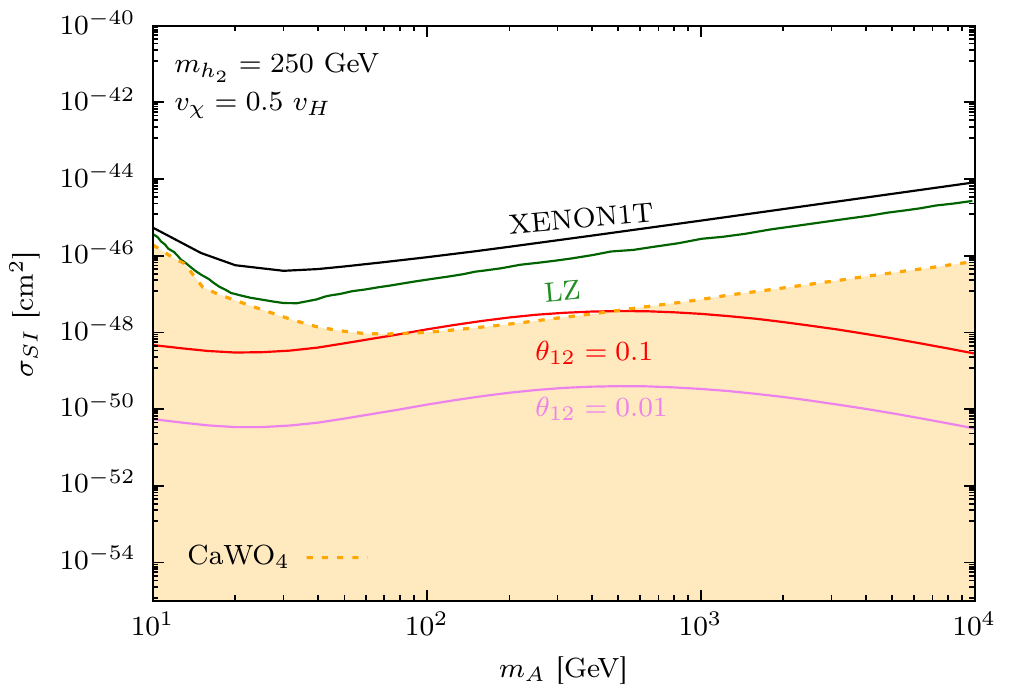}
\includegraphics[height=4.5cm, width=5.8cm]{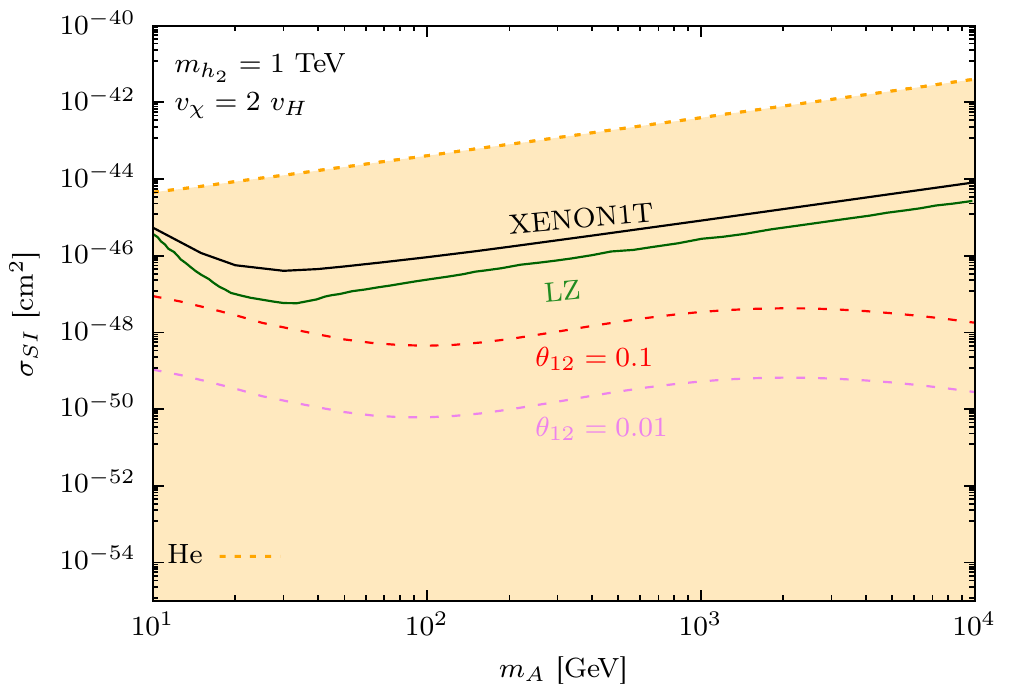}
\includegraphics[height=4.5cm, width=5.8cm]{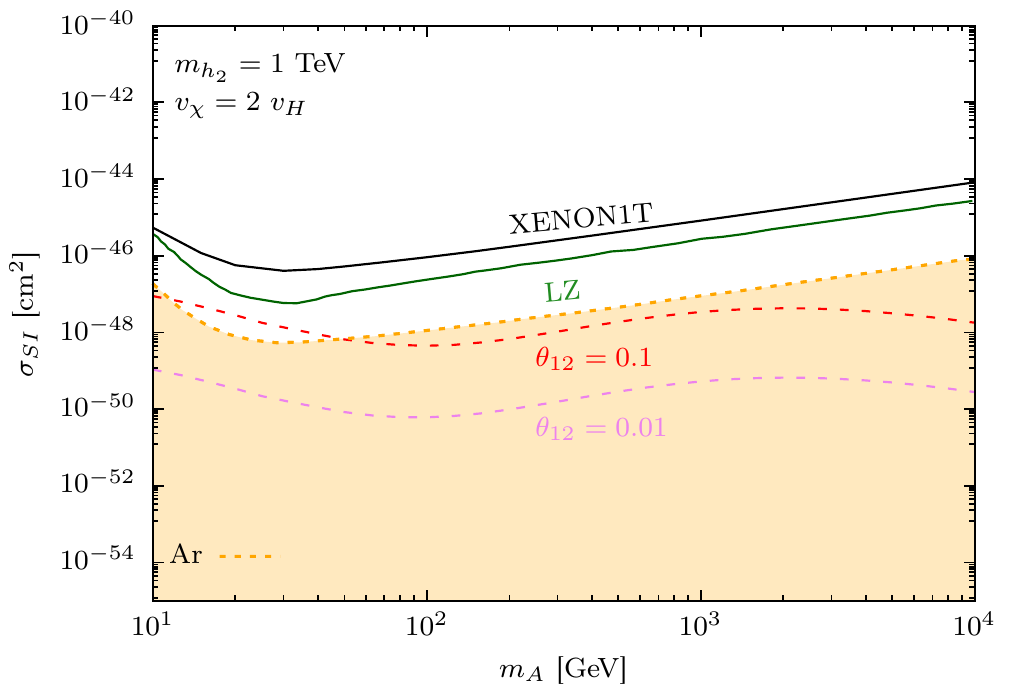}
\includegraphics[height=4.5cm, width=5.8cm]{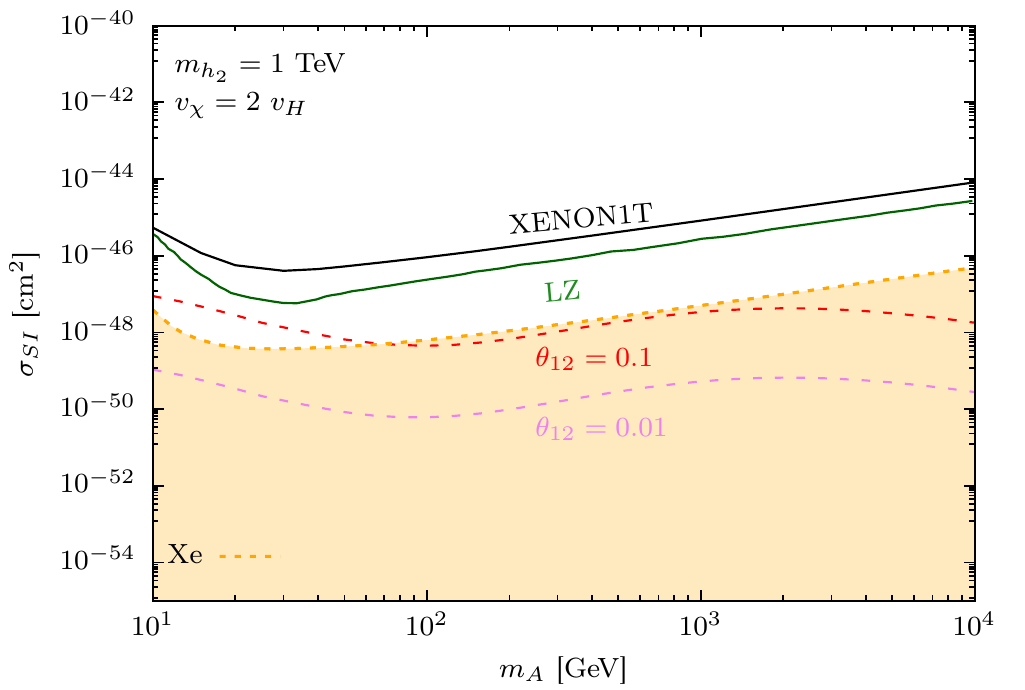}
\includegraphics[height=4.5cm, width=5.8cm]{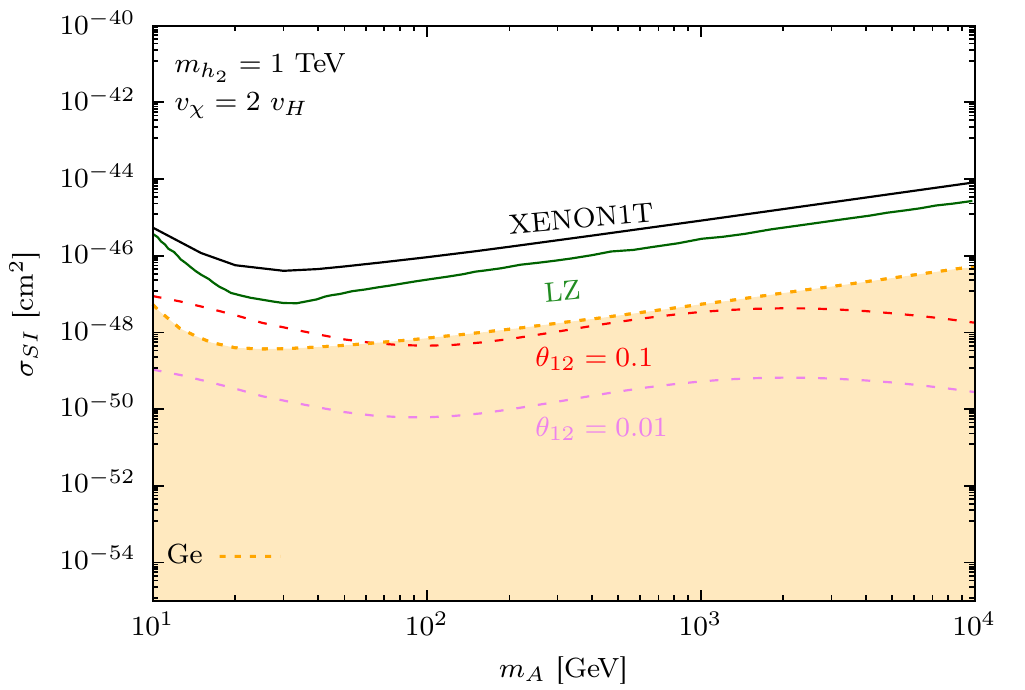}
\includegraphics[height=4.5cm, width=5.8cm]{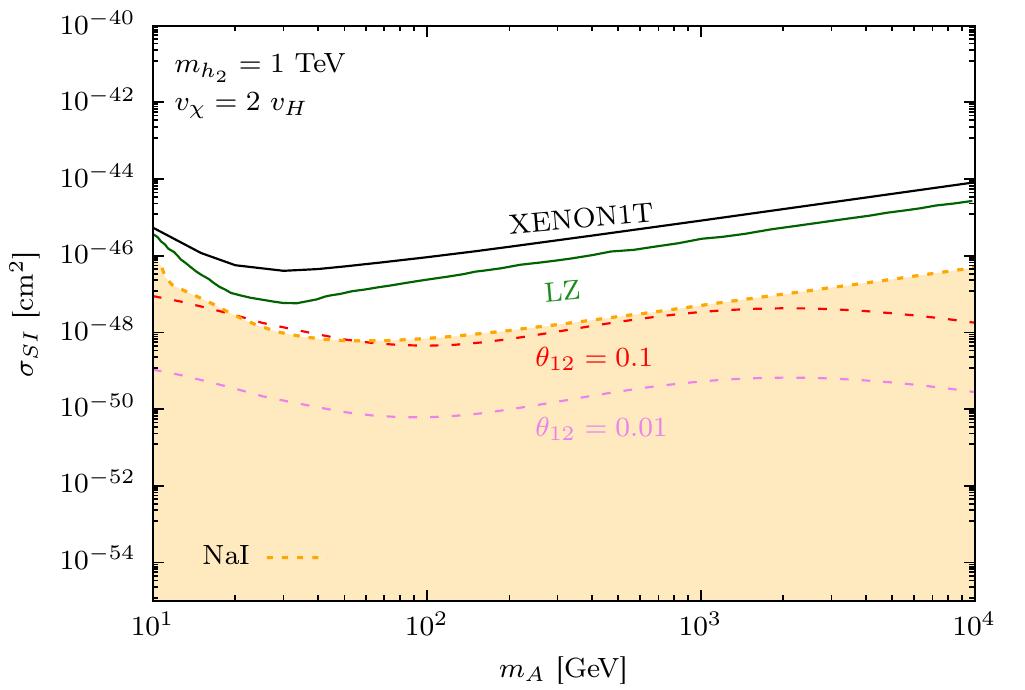}
\includegraphics[height=4.5cm, width=5.8cm]{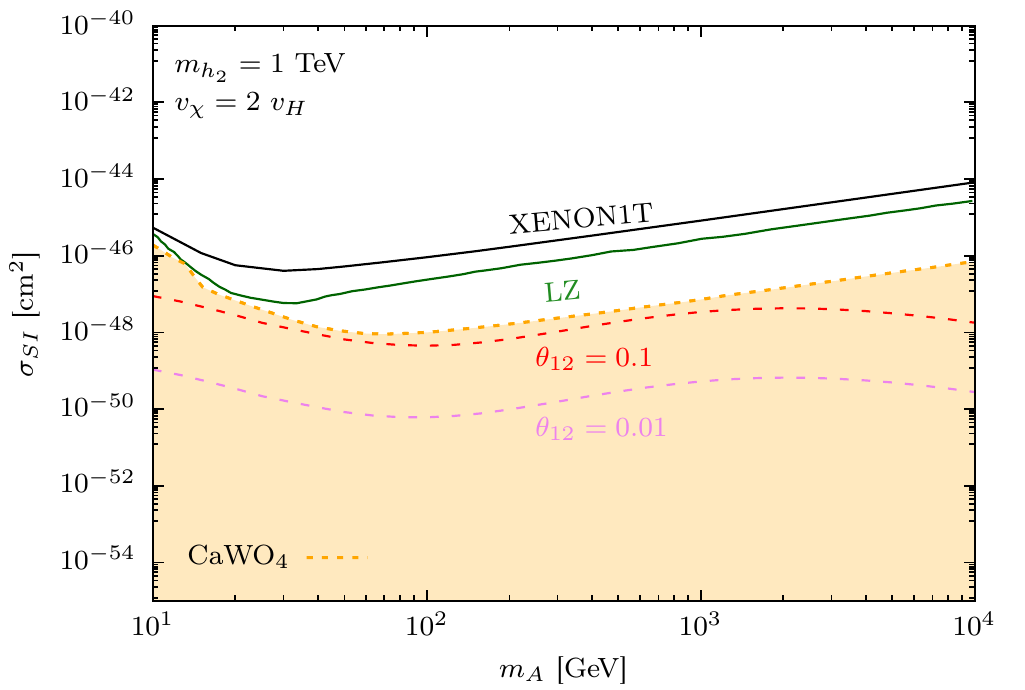}
\caption{We have fixed $m_{h_2}=250~(1000)$ GeV and $v_{\chi}=0.5~(2)~v_H$ in upper~(remaining) two panels. The red and pink curves in each plot are the DM-nucleon spin-independent cross section~($\sigma_{SI}$) for scalar mixing $\theta_{12}=0.1,0.01$ respectively. The black and dark-green curves are the direct detection limit from XENON1T~\cite{Aprile:2018dbl} and LZ~\cite{Akerib:2018lyp} experiments. The orange region is neutrino floor data~\cite{O_Hare_2021} for various targets such as Helium~(He), Argon~(Ar), Xenon~(Xe), Germanium~(Ge), NaI, CaWO$_4$ as labelled in each plot.}  
\label{dd}
\end{figure}

The main advantage of having a pseudo scalar DM is that the DM-nucleon scattering cross section vanishes in the non-relativistic limit which we need to check that this is the case in our model too.

The relevant interaction vertices for scattering matrix is A$-$A$-h_i$, which simplify in the limit~$v_{\phi} >> v_H,~v_{\chi}$ and given as follow,

\begin{align*}
g_{A A h_1} \approx v_H \lambda_{H\chi} \cos\theta_{12} + 
  2. v_{\chi} \lambda_{\chi} \sin\theta_{12}, \hskip 0.5cm  
g_{A A h_2} \approx - v_H \lambda_{H\chi} \sin\theta_{12} + 
  2. v_{\chi} \lambda_{\chi} \cos\theta_{12}, \ \hskip 0.3cm
 g_{A A h_3} \approx 0
\end{align*} 

Hence, the scattering matrix is some of only two Feynman diagrams mediated by the two lighter higgs $h_1,~h_2$.

\begin{align*}
\mathcal{M} \sim \frac{g_{A A h_1}g_{h_1 f \bar{f}}}{q-m^2_{h_1}} + \frac{g_{A A h_2}g_{h_2 f \bar{f}}}{q-m^2_{h_2}} 
\end{align*}

here $g_{h_i f \bar{f}}$ is the coupling for Higgs-SM fermions interaction and q is the momentum transfer. One can see by using equation~\ref{eq:quartic} that the tree label amplitude vanishes in the non-relativistic limit as it is shown for the simpler case in~\cite{Gross_2017}. However, one loop contribution could be finite and it has been studied in ref~\cite{Ishiwata_2018}. There will be three types of Feynman diagrams, namely, 1. Self-energy, 2. Vertex corrections, 3. Box and triangle diagrams, as shown in ref~\cite{Ishiwata_2018}, contribute to one loop. The generic expression for the scattering cross-section, which is the sum of all these Feynman diagrams, is given by,
\begin{align}
\sigma^N_{\text{SI}}= \frac{1}{\pi}\left(\frac{m_N}{m_A+m_N}\right)^2 \vert f^N_{\text{scalar}}+f^N_{\text{twist2}}\vert^2
\label{sigmalo}
\end{align}
here $m_N$ is mass of nucleon and definition of functions $f^N_{\text{scalar}}, ~f^N_{\text{twist2}}$ is given in ref~\cite{Ishiwata_2018}. 

\begin{figure}[h!]
\includegraphics[height=5.5cm, width=7.5cm]{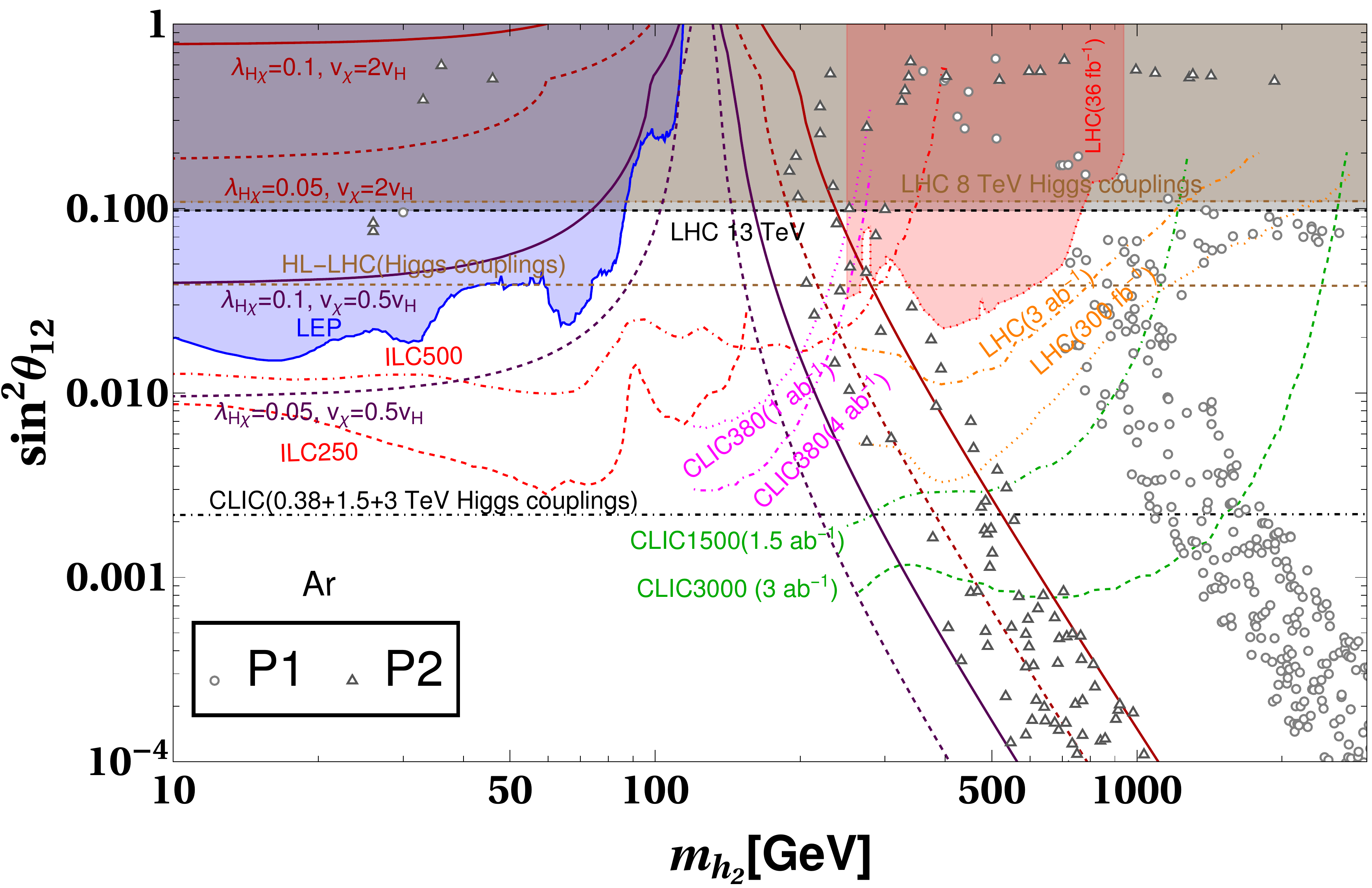}
\includegraphics[height=5.5cm, width=7.5cm]{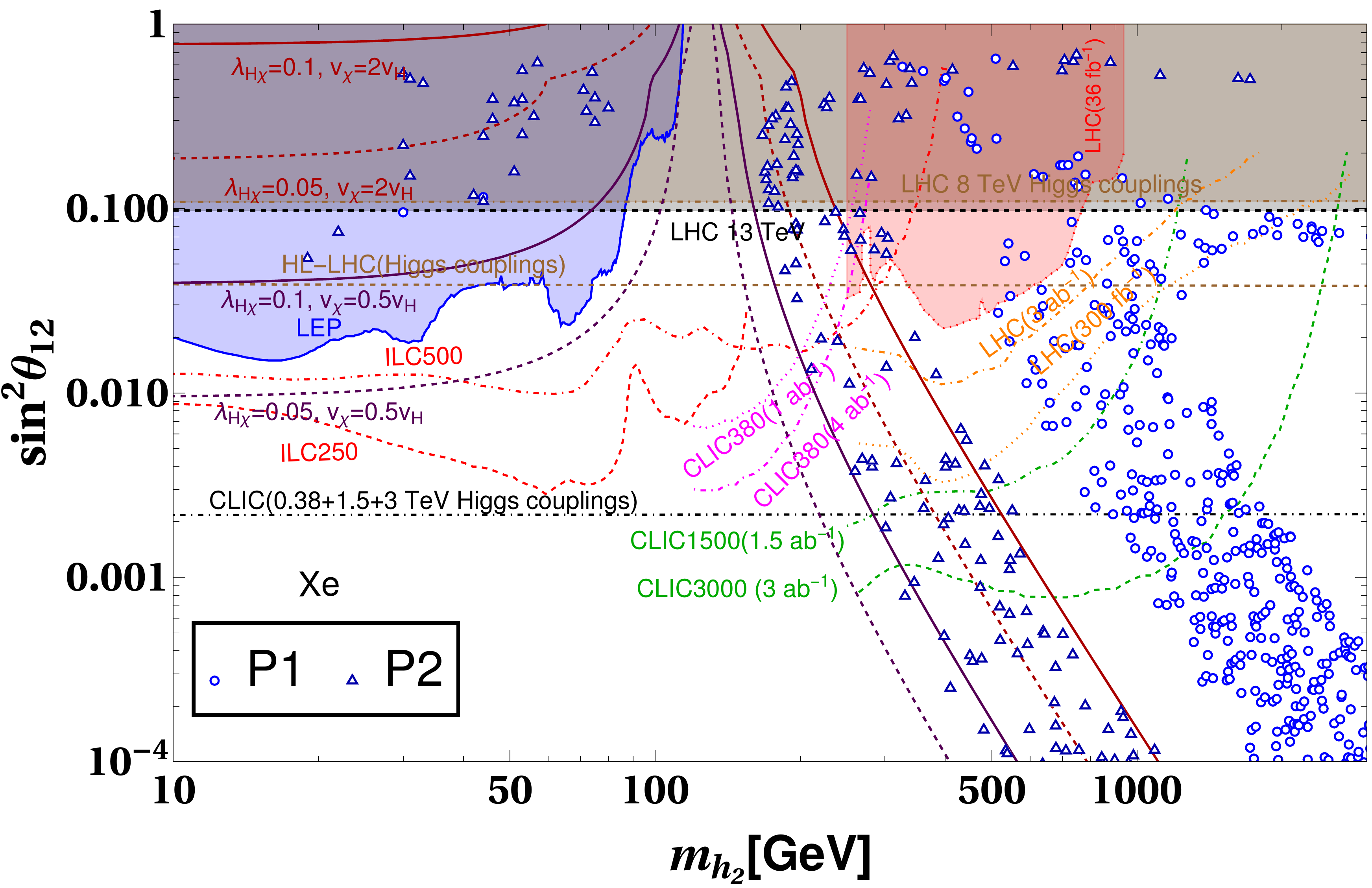}
\includegraphics[height=5.5cm, width=7.5cm]{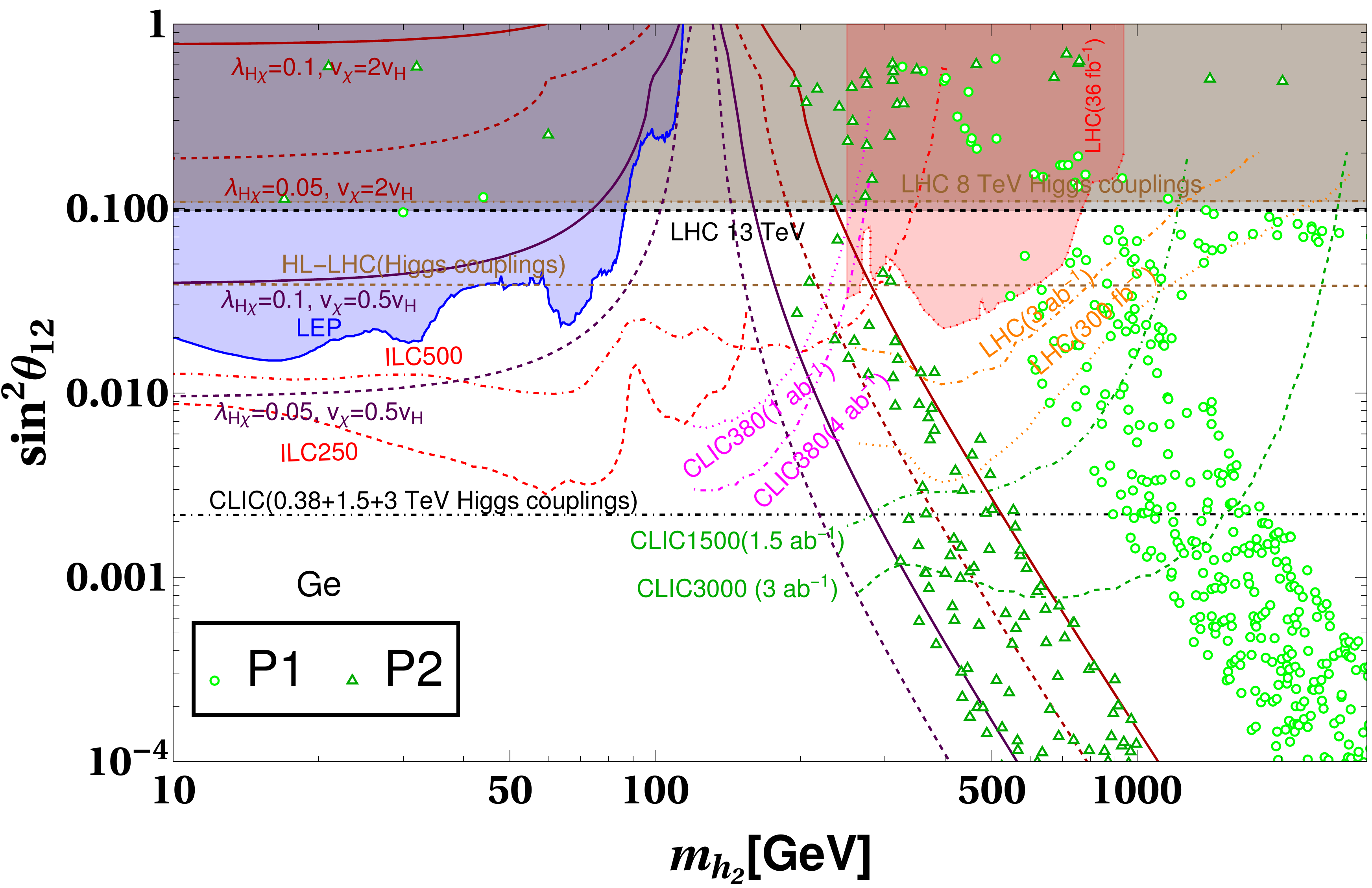}
\includegraphics[height=5.5cm, width=7.5cm]{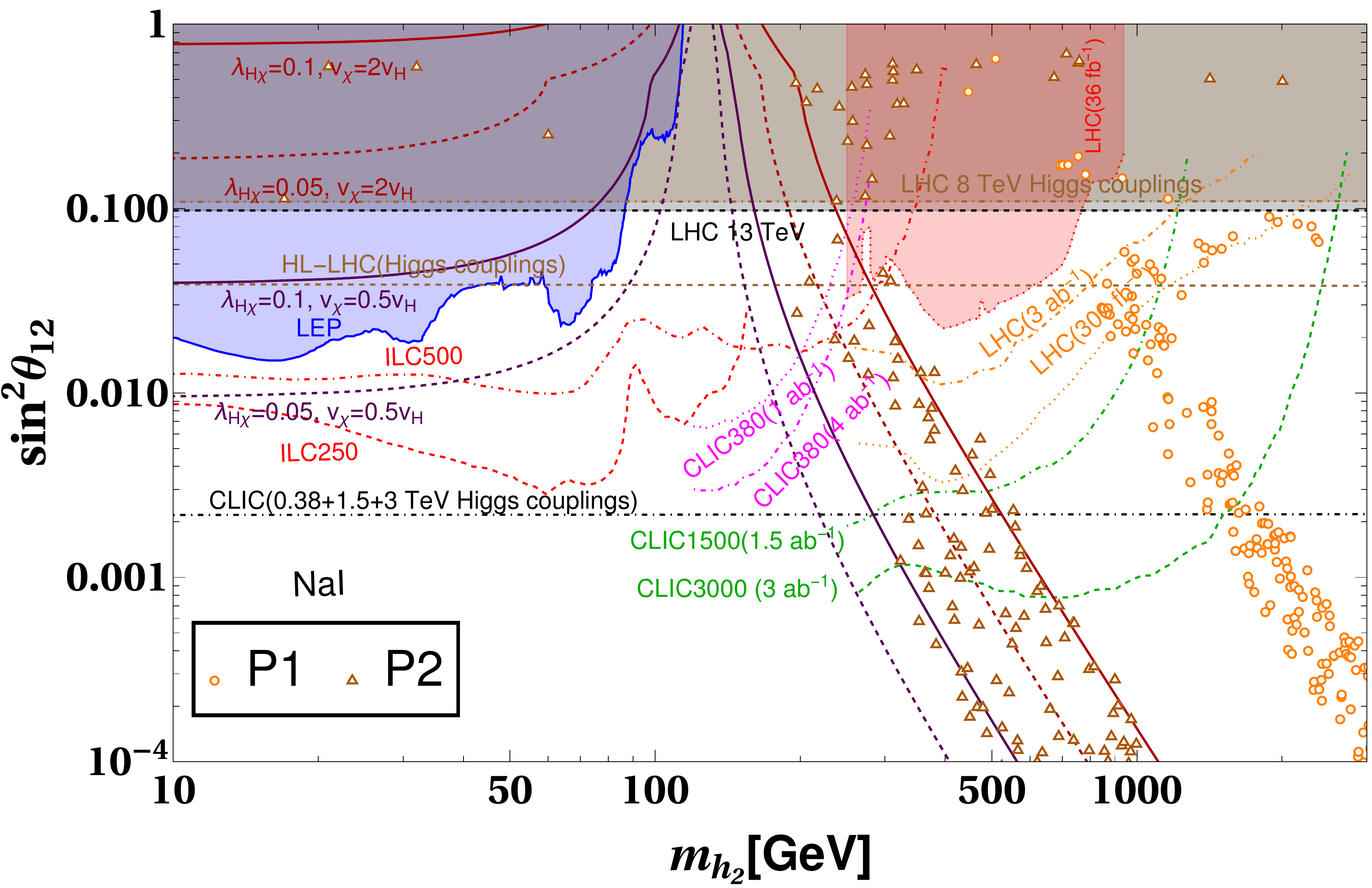}
\includegraphics[height=5.5cm, width=7.5cm]{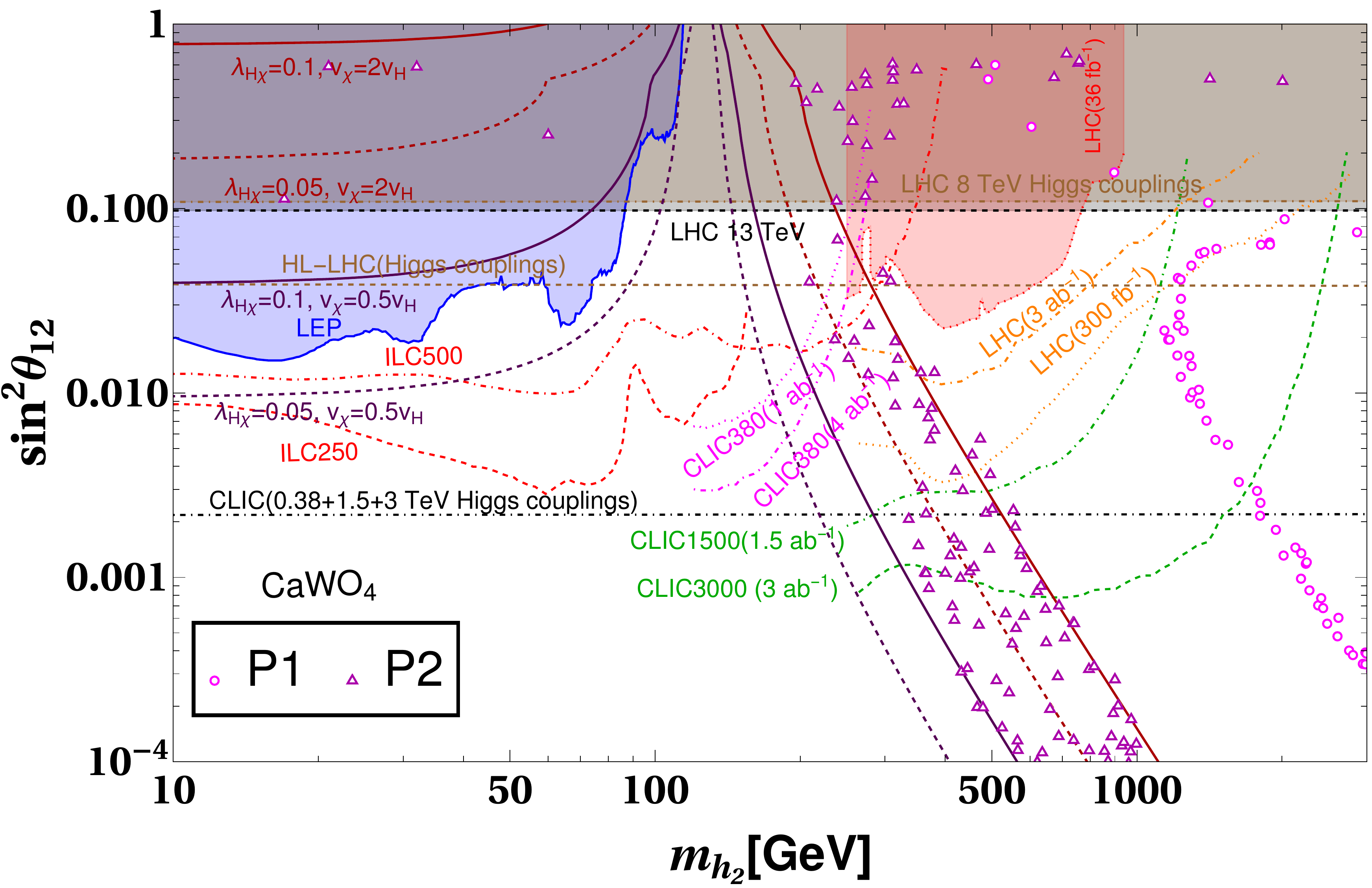}
\vspace{2mm}
\caption{The plot here is adapted from~\cite{Das:2022oyx}. Several bounds from colliders such as LHC \cite{deBlas:2018mhx}, LEP\cite{LEPWorkingGroupforHiggsbosonsearches:2003ing}, prospective colliders such as ILC \cite{Wang:2020lkq} and CLIC \cite{deBlas:2018mhx} on the scalar mixing $\sin^2\theta_{12}$ as a function of $m_{h_2}$ is shown here. Shaded regions are ruled out by corresponding experiments. Limits on the scalar mixing from Eq.~\ref{eq:quartic} are also shown for different $\lambda_{H\chi}=$0.1, 0.05 respectively. Constraints from direct detection for $(m_A, v_{\chi}) =$ $ P1\sim(20~\text{GeV}, ~2v_H),~ P2\sim(200~\text{GeV}, ~0.5v_H)$ is shown when the neutrino floor is due to Ar, Xe, Ge, NaI and CaWO$_4$. }  
\label{sinmh2}
\end{figure}

In our model, the third Higgs $h_3$ and Z$'$ are too heavy, to play any significant role in the scattering matrix calculation, therefore, It is straightforward to follow the formalism developed in the reference to evaluate the scattering cross section at one loop level using equation \ref{sigmalo} in our model.     

In fig~\ref{dd}, we check for the direct detection constraint on $(\sigma_{\text{SI}}, m_A)$ plane. We have fixed $m_{h_2}=250$ GeV and $v_{\chi}=0.5~v_H$ for the first two panels and for the remaining two panels, we keep $m_{h_2}=1$ TeV and $v_{\chi}=2 ~v_H$. The red and blue curve in each plot is the prediction from the model for the $h_1-h_2$ mixing angle $\theta_{12}=0.1,~0.01$ respectively. The Black and dark-green curve is the limit from XENON1T~\cite{Aprile:2018dbl} and LZ~\cite{Akerib:2018lyp}. The light orange region is the neutrino background in each plot due to six different targets Helium~(He), Argon~(Ar), Xenon~(Xe), Germanium~(Ge), NaI and CaWO$_4$. The neutrino floor data is taken from ~\cite{O_Hare_2021}. One can see the theoretical prediction e.g. $\theta_{12}=0.1$ is above the neutrino floor around $m_A=200, 20$ GeV when $v_{\chi} = 0.5v_H, ~2v_H$ respectively and the theoretical value is comparable to the LZ and XENON1T data for most of the $m_A$ range. The neutrino floor limit from six different target materials suggests an improved experiment setup to search for the direct detection signals. The direct detection signals can be searched in the allowed parameter space of our model as shown in fig.~\ref{dd}.

\subsection{Constraining $\sin{\theta_{12}}-m_{h_2}$ plane}
\label{theta_mh2}

In this section, we constraint the ($\sin\theta_{12}$, $m_{h_2}$) plane using direct detection study done in fig~\ref{dd}. To do this, we calculate the DM-nucleon cross section using equation~\ref{sigmalo} for $(m_A, v_{\chi}) = P1\sim(20~\text{GeV}, ~2 v_H),~ P2\sim(200~\text{GeV}, ~0.5 v_H)$, which is allowed points for $\theta_{12}=0.1$ by bounds as shown in fig~\ref{dd}. We then constrained the scattering cross section using limits from LZ and neutrino floors data for the corresponding points while taking $\theta_{12}$ and $m_{h_2}$ as free parameter. Since, neutrino floor data is from five different targets such as Ar, Xe, Ge, NaI and CaWO$_4$, therefore, we have five distinct plots for scalar mixing as shown in fig~\ref{sinmh2}. We do not consider neutrino floor data from the Helium target, since it gives a much larger cross-section as seen in figure~\ref{dd}, thus, He is not a good choice for direct searches and hence, we do not consider it for further analysis. 

There are several colliders bounds such as LHC \cite{deBlas:2018mhx}, LEP \cite{LEPWorkingGroupforHiggsbosonsearches:2003ing}, prospective colliders e.g. ILC \cite{Wang:2020lkq} and CLIC \cite{deBlas:2018mhx} are adapted here from the reference~\cite{Das:2022oyx}. The LEP bounds shown by the blue solid line consider both $Z\to \mu^+\mu^-$ and $e^+e^-$ modes. Bounds from LHC and High Luminosity LHC (HL-LHC) at the 8 TeV are shown by the brown dot-dashed and dashed lines respectively~\cite{deBlas:2018mhx}. LHC bounds at 13 TeV are shown by the black dashed line from the ATLAS results \cite{ATLAS:2020qdt}. The prospective bounds estimated form $h_2 \to ZZ$ mode at the LHC at 300 (3000) fb$^{-1}$ luminosity using orange dot dashed (dotted) line from \cite{Buttazzo:2018qqp}. The prospective bounds from the combined CLIC analyses art 380 GeV, 1.5 TeV and 3 TeV are shown by the black dot dashed line. 

We also superimpose the limits on the scalar mixing~$\theta_{12}$ using Eq.~\ref{eq:quartic} by considering two different choices of $\lambda_{H\chi}$ as $0.1$, $0.05$ for each point P1 and P2. The scalar mixing $\sin^2\theta_{12} < 0.001$ is respected by each constraint as seen in fig~\ref{sinmh2}.

\section{Conclusion}
\label{sec:conclusion}

We have considered a generic $U(1)_X$ extension of SM which accommodates a pseudo scalar DM candidate and neutrino mass generation mechanism. In the model, we have three RHNs and two complex scalars $\Phi$, $\chi$ additionally and all are charged under the $U(1)_X$ gauge group. $U(1)_X$ gauge is broken by the vev of $\Phi$. $\chi$ field also gets vev $v_{\chi}$ which results in a massive pseudo scalar particle. The pseudo scalar has coupling with Higgs and neutral gauge bosons and its  phenomenology is controlled by a few parameters such as $x_H$, $\kappa$, $m_{h_i}$, $g'_{1}$, $M_{Z'}$ and $m_A$. We have found that pseudo scalar have sufficient parameter space which is satisfied by DM lifetime constraint while remaining consistent. We found the allowed parameter space by the relic density constraints, Invisible Higgs width constraint and other bounds. Direct detection check is also done in our model. The tree-level DM-nucleon scattering amplitude vanishes in non-relativistic limit but one loop contribution is finite. We have shown the one-loop contribution to scattering cross-section against the experimental limit from XENON1T, LZ and neutrino floors from five different targets. In our model, the pseudo scalar DM have sufficient parameter space which is consistent with several theoretical and experimental constraints and can be tested at future colliders and direct detection experiments. 

\begin{acknowledgements}

I am much obliged to Dr$.$ Arindam Das and Dr$.$ Sanjoy Mandal for their invaluable time for discussions and suggestions. 
\end{acknowledgements}

\appendix

\section{Anomaly cancellations}
\label{anomaly}

The gauge and mixed gauge-gravity anomaly cancellation conditions on the $U(1)_X$ charges are as follows:

\begin{align*}
{U}(1)_X \otimes \left[ {SU}(3)_c \right]^2&\ :&
			2x_q - x_u - x_d &\ =\  0, \nonumber \\
{U}(1)_X \otimes \left[ {SU}(2)_L \right]^2&\ :&
			3x_q + x_\ell &\ =\  0, \nonumber \\
{U}(1)_X \otimes \left[ {U}(1)_Y \right]^2&\ :&
			x_q - 8 x_u - 2x_d + 3x_\ell - 6x_e &\ =\  0, \nonumber \\
\left[ {U}(1)_X \right]^2 \otimes {U}(1)_Y&\ :&
			{x_q}^2 - {2x_u}^2 + {x_d}^2 - {x_\ell}^2 + {x_e}^2 &\ =\  0, \nonumber \\
\left[ {U}(1)_X \right]^3&\ :&
			{6x_q}^3 - {3x_u}^3 - {3x_d}^3 + {2x_\ell}^3 - {x_\nu}^3 - {x_e}^3 &\ =\  0, \nonumber \\
{U}(1)_X \otimes \left[ {\rm grav.} \right]^2&\ :&
			6x_q - 3x_u - 3x_d + 2x_\ell - x_\nu - x_e &\ =\  0. 
\label{anom-f}
\end{align*}

The general $U(1)_X$ charge assignment is the linear combination of the $U(1)_Y$ and $B-L$ charges, as shown in Table~\ref{tab1}.

\section{Decay widths}
\label{dwidths}
The relevant vertices and decay width expressions are as follows,
\begin{align}
 \mathcal{L}_{A Z h_i} =& \sum_i \lambda_{AZh_i}  Z_\mu ( h_i \partial^\mu A - A \partial^\mu h_i),
 \\
 \mathcal{L}_{AZ'h_i} =& \sum_i \lambda_{AZ'h_i}   Z'_\mu (h_i \partial^\mu A - A \partial^\mu h_i).
\end{align}
where,
\begin{align}
\lambda_{AZh_i}=\frac{g_X x_\Phi \sin\theta}{\sqrt{1-\kappa^2}}(2\cos\theta_\eta O_{R_{i3}} -\sin\theta_\eta O_{R_{i2}}),\,\,\lambda_{AZ'h_i}=\frac{g_X x_\Phi \cos\theta}{\sqrt{1-\kappa^2}}(2\cos\theta_\eta O_{R_{i3}} -\sin\theta_\eta O_{R_{i2}})
\end{align}

The couplings between the gauge bosons $Z, Z'$ and the (axial) vector currents of the SM fermion $f$ is defined by
\begin{align}
 \mathcal{L}_{Z' \bar{f}f}= Z'_{\mu} \bar{f} \gamma^\mu \Bigl[ g^{Z'f}_{L}P_L +g^{Z'f}_{R} P_R \Bigl] f,\,\, \mathcal{L}_{Z \bar{f}f}= Z_{\mu} \bar{f} \gamma^\mu \Bigl[ g^{Zf}_{L}P_L +g^{Zf}_{R} P_R \Bigl] f,
\end{align}
where
\begin{align}
& g^{Z\ell}_{R}=-\frac{1}{\sqrt{1-\kappa^2}}\Big(g_X \sin\theta \big(x_H+x_\Phi\big)-g_Z \big(\sin^2\theta_W\cos\theta \sqrt{1-\kappa^2}+\kappa \sin\theta \sin\theta_W\big)\Big),\nonumber\\
& g^{Z\ell}_{L}=-\frac{1}{2\sqrt{1-\kappa^2}}\Big(g_X \sin\theta \big(x_H+2x_\Phi\big)-g_Z \big(\kappa \sin\theta \sin\theta_W-(1-2\sin^2\theta_W)\cos\theta \sqrt{1-\kappa^2}\big)\Big),\nonumber\\
& g^{Zu}_{R}=\frac{1}{3\sqrt{1-\kappa^2}}\Big(g_X \sin\theta \big(2x_H+x_\Phi\big)-2g_Z\big(\cos\theta \sin^2\theta_W\sqrt{1-\kappa^2}+\kappa \sin\theta\sin\theta_W\big)\Big),\nonumber\\
& g^{Zu}_{L}=\frac{1}{6\sqrt{1-\kappa^2}}\Big(g_X\sin\theta \big(x_H+2x_\Phi\big)-g_Z\big(\cos\theta \sqrt{1-\kappa^2}(4\sin^2\theta_W-3)+\kappa\sin\theta\sin\theta_W\big)\Big),\nonumber\\
& g^{Zd}_{R}=\frac{1}{3\sqrt{1-\kappa^2}}\Big(g_X\sin\theta\big(x_\Phi-x_H\big)+g_Z\big(\cos\theta\sin^2\theta_W\sqrt{1-\kappa^2}+\kappa \sin\theta\sin\theta_W\big)\Big),\nonumber\\
& g^{Zd}_{L}=\frac{1}{6\sqrt{1-\kappa^2}}\Big(g_X\sin\theta \big(x_H+2x_\Phi\big)-g_Z\big(\cos\theta\sqrt{1-\kappa^2}(3-2\sin^2\theta_W)+\kappa\sin\theta\sin\theta_W\big)\Big)\nonumber
\end{align}
where $g^{Z'\ell/u/d}_{L/R}=g^{Z\ell/u/d}_{L/R}\Big(\sin\theta\to \cos\theta; \cos\theta\to -\sin\theta \Big)$.

\begin{align}
& \Gamma(A\to h_i X)=\frac{|\lambda_{AXh_i}|^2 m_A^3}{16\pi M_{X}^2}\lambda^{\frac{3}{2}}\Big(1,\frac{M_X^2}{m_A^2},\frac{m_{h_i}^2}{m_A^2}\Big),\\
& \Gamma(A\to N_i N_i)=\frac{\cos^2\theta_\eta}{8\pi v_\Phi^2}m_A M_{N_i}^2 \sqrt{1-4\frac{M_{N_i}^2}{m_A^2}}
\end{align}
$\lambda(a,b,c) = a^2 + b^2 + c^2 - 2 ab - 2 bc - 2 ca $.


\begin{align}
& d\Gamma(A\to h_i f\bar{f})=\frac{m_A^5}{384 \pi^3}\Big[\big(|A_L^{if}(z)|^2+|A_R^{if}(z)|^2\big) \big\{\lambda\big(1,x_{h_i}^2,z\big)+\frac{x_f^2}{z}\big(2(1-x_{h_i}^2)^2+2(1+x_{h_i}^2)z-z^2\big)\big\}\nonumber \\
& -6\text{Re}\big(A_L^{if*}(z) A_R^{if}(z)\big) x_f^2 (2(1+x_{h_i}^2)-z) \Big]\frac{dz}{z}\lambda^{\frac{1}{2}}\big(1,x_{h_i}^2,z\big) \lambda^{\frac{1}{2}}\big(1,x_{f}^2,x_f^2\big);\,\,\, 4 x_f^2 \leq z \leq (1-x_{h_i})^2,
\end{align}
where,
\begin{align}
A_{L(R)}^{if}(z)=\frac{\lambda_{AZh_i} g_{L(R)}^{Zf}}{z m_A^2 - M_Z^2 + i \Gamma_Z M_Z}+\frac{\lambda_{AZ'h_i} g_{L(R)}^{Z'f}}{z m_A^2 - M_Z^{'2} + i \Gamma_{Z'} M_{Z'}}
\end{align}
\begin{align}
\Gamma(A\to X f\bar{f})=\frac{|H_X(z)|^2 m^9 x_f^2}{128\pi^3 M_X^2 v_H^2} \big(z-4x_f^2\big)\lambda^{\frac{3}{2}}\big(1,\frac{M_X^2}{m_A^2},z\big)\frac{dz}{z} \lambda^{\frac{1}{2}}\big(1,x_{f}^2,x_f^2\big)
\end{align}
\begin{align}
H_Z(z)=\sum_{i=1}^3 \frac{\lambda_{AZh_i} \mathcal{O}_{R_{i1}}}{z m_A^2-m_{h_i}^2+i\Gamma_{h_i}m_{h_i}},\,\, H_{Z'}(z)=\sum_{i=1}^3 \frac{\lambda_{AZ'h_i} \mathcal{O}_{R_{i1}}}{z m_A^2-m_{h_i}^2+i\Gamma_{h_i}m_{h_i}}
\end{align}

\bibliographystyle{utphys}
\bibliography{bibitem} 

\providecommand{\href}[2]{#2}\begingroup\raggedright\begin{thebibliography}{10}

\bibitem{Garrett_2011}
K.~Garrett and G.~Duda, ``Dark matter: A primer,''
  \href{http://dx.doi.org/10.1155/2011/968283}{{\em Advances in Astronomy}
  {\bfseries 2011} (2011) 1--22}.
  \url{https://doi.org/10.1155%2F2011%2F968283}.

\bibitem{Profumo:2009tb}
S.~Profumo, K.~Sigurdson, and L.~Ubaldi, ``{Can we discover multi-component
  WIMP dark matter?},''
  \href{http://dx.doi.org/10.1088/1475-7516/2009/12/016}{{\em JCAP} {\bfseries
  12} (2009) 016}, \href{http://arxiv.org/abs/0907.4374}{{\ttfamily
  arXiv:0907.4374 [hep-ph]}}.

\bibitem{Bertone:2004pz}
G.~Bertone, D.~Hooper, and J.~Silk, ``{Particle dark matter: Evidence,
  candidates and constraints},''
  \href{http://dx.doi.org/10.1016/j.physrep.2004.08.031}{{\em Phys. Rept.}
  {\bfseries 405} (2005) 279--390},
  \href{http://arxiv.org/abs/hep-ph/0404175}{{\ttfamily arXiv:hep-ph/0404175}}.

\bibitem{Bartelmann:1999yn}
M.~Bartelmann and P.~Schneider, ``{Weak gravitational lensing},''
  \href{http://dx.doi.org/10.1016/S0370-1573(00)00082-X}{{\em Phys. Rept.}
  {\bfseries 340} (2001) 291--472},
  \href{http://arxiv.org/abs/astro-ph/9912508}{{\ttfamily
  arXiv:astro-ph/9912508}}.

\bibitem{Clowe:2003tk}
D.~Clowe, A.~Gonzalez, and M.~Markevitch, ``{Weak lensing mass reconstruction
  of the interacting cluster 1E0657-558: Direct evidence for the existence of
  dark matter},'' \href{http://dx.doi.org/10.1086/381970}{{\em Astrophys. J.}
  {\bfseries 604} (2004) 596--603},
  \href{http://arxiv.org/abs/astro-ph/0312273}{{\ttfamily
  arXiv:astro-ph/0312273}}.

\bibitem{Harvey:2015hha}
D.~Harvey, R.~Massey, T.~Kitching, A.~Taylor, and E.~Tittley, ``{The
  non-gravitational interactions of dark matter in colliding galaxy
  clusters},'' \href{http://dx.doi.org/10.1126/science.1261381}{{\em Science}
  {\bfseries 347} (2015) 1462--1465},
  \href{http://arxiv.org/abs/1503.07675}{{\ttfamily arXiv:1503.07675
  [astro-ph.CO]}}.

\bibitem{Hinshaw:2012aka}
{\bfseries WMAP} Collaboration, G.~Hinshaw {\em et~al.}, ``{Nine-Year Wilkinson
  Microwave Anisotropy Probe (WMAP) Observations: Cosmological Parameter
  Results},'' \href{http://dx.doi.org/10.1088/0067-0049/208/2/19}{{\em
  Astrophys. J. Suppl.} {\bfseries 208} (2013) 19},
  \href{http://arxiv.org/abs/1212.5226}{{\ttfamily arXiv:1212.5226
  [astro-ph.CO]}}.

\bibitem{Ade:2015xua}
{\bfseries Planck} Collaboration, P.~A.~R. Ade {\em et~al.}, ``{Planck 2015
  results. XIII. Cosmological parameters},''
  \href{http://dx.doi.org/10.1051/0004-6361/201525830}{{\em Astron. Astrophys.}
  {\bfseries 594} (2016) A13},
  \href{http://arxiv.org/abs/1502.01589}{{\ttfamily arXiv:1502.01589
  [astro-ph.CO]}}.

\bibitem{Schumann_2019}
M.~Schumann, ``Direct detection of {WIMP} dark matter: concepts and status,''
  \href{http://dx.doi.org/10.1088/1361-6471/ab2ea5}{{\em Journal of Physics G:
  Nuclear and Particle Physics} {\bfseries 46} no.~10, (Aug, 2019) 103003}.
  \url{https://doi.org/10.1088%2F1361-6471%2Fab2ea5}.

\bibitem{Lisanti:2016jxe}
M.~Lisanti, \href{http://dx.doi.org/10.1142/9789813149441_0007}{``{Lectures on
  Dark Matter Physics},''} in {\em {Theoretical Advanced Study Institute in
  Elementary Particle Physics}: {New Frontiers in Fields and Strings}},
  pp.~399--446.
\newblock 2017.
\newblock \href{http://arxiv.org/abs/1603.03797}{{\ttfamily arXiv:1603.03797
  [hep-ph]}}.

\bibitem{Gross_2017}
C.~Gross, O.~Lebedev, and T.~Toma, ``Cancellation mechanism for
  dark-matter{\textendash}nucleon interaction,''
  \href{http://dx.doi.org/10.1103/physrevlett.119.191801}{{\em Physical Review
  Letters} {\bfseries 119} no.~19, (Nov, 2017) }.
  \url{https://doi.org/10.1103%2Fphysrevlett.119.191801}.

\bibitem{Abe_2020}
Y.~Abe, T.~Toma, and K.~Tsumura, ``Pseudo-nambu-goldstone dark matter from
  gauged u(1)b-l symmetry,''
  \href{http://dx.doi.org/10.1007/jhep05(2020)057}{{\em Journal of High Energy
  Physics} {\bfseries 2020} no.~5, (May, 2020) }.
  \url{https://doi.org/10.1007%2Fjhep05%282020%29057}.

\bibitem{Abe_2021}
Y.~Abe, T.~Toma, K.~Tsumura, and N.~Yamatsu, ``Pseudo-nambu-goldstone dark
  matter model inspired by grand unification,''
  \href{http://dx.doi.org/10.1103/physrevd.104.035011}{{\em Physical Review D}
  {\bfseries 104} no.~3, (Aug, 2021) }.
  \url{https://doi.org/10.1103%2Fphysrevd.104.035011}.

\bibitem{Gola:2021abm}
S.~Gola, S.~Mandal, and N.~Sinha, ``{ALP-portal majorana dark matter},''
  \href{http://dx.doi.org/10.1142/S0217751X22501317}{{\em Int. J. Mod. Phys. A}
  {\bfseries 37} no.~22, (2022) 2250131},
  \href{http://arxiv.org/abs/2106.00547}{{\ttfamily arXiv:2106.00547
  [hep-ph]}}.

\bibitem{Okada_2021}
N.~Okada, D.~Raut, and Q.~Shafi, ``Pseudo-goldstone dark matter in a gauged
  $b-l$ extended standard model,''
  \href{http://dx.doi.org/10.1103/physrevd.103.055024}{{\em Physical Review D}
  {\bfseries 103} no.~5, (Mar, 2021) }.
  \url{https://doi.org/10.1103%2Fphysrevd.103.055024}.

\bibitem{Oda_2015}
S.~Oda, N.~Okada, and D.~suke Takahashi, ``Classically conformal u(1)$'$
  extended standard model and higgs vacuum stability,''
  \href{http://dx.doi.org/10.1103/physrevd.92.015026}{{\em Physical Review D}
  {\bfseries 92} no.~1, (Jul, 2015) }.
  \url{https://doi.org/10.1103%2Fphysrevd.92.015026}.

\bibitem{Das_2019}
A.~Das, N.~Okada, S.~Okada, and D.~Raut, ``Probing the seesaw mechanism at the
  250 {GeV} {ILC},''
  \href{http://dx.doi.org/10.1016/j.physletb.2019.134849}{{\em Physics Letters
  B} {\bfseries 797} (Oct, 2019) 134849}.
  \url{https://doi.org/10.1016%2Fj.physletb.2019.134849}.

\bibitem{Das_2022}
A.~Das, S.~Mandal, T.~Nomura, and S.~Shil, ``Heavy majorana neutrino pair
  production from z` at hadron and lepton colliders,''
  \href{http://dx.doi.org/10.1103/physrevd.105.095031}{{\em Physical Review D}
  {\bfseries 105} no.~9, (May, 2022) }.
  \url{https://doi.org/10.1103%2Fphysrevd.105.095031}.

\bibitem{Das:2016zue}
A.~Das, S.~Oda, N.~Okada, and D.-s. Takahashi, ``{Classically conformal U(1)'
  extended standard model, electroweak vacuum stability, and LHC Run-2
  bounds},'' \href{http://dx.doi.org/10.1103/PhysRevD.93.115038}{{\em Phys.
  Rev. D} {\bfseries 93} no.~11, (2016) 115038},
  \href{http://arxiv.org/abs/1605.01157}{{\ttfamily arXiv:1605.01157
  [hep-ph]}}.

\bibitem{Okada:2016tci}
N.~Okada and S.~Okada, ``{$Z^\prime$-portal right-handed neutrino dark matter
  in the minimal U(1)$_X$ extended Standard Model},''
  \href{http://dx.doi.org/10.1103/PhysRevD.95.035025}{{\em Phys. Rev. D}
  {\bfseries 95} no.~3, (2017) 035025},
  \href{http://arxiv.org/abs/1611.02672}{{\ttfamily arXiv:1611.02672
  [hep-ph]}}.

\bibitem{Bandyopadhyay:2017bgh}
P.~Bandyopadhyay, E.~J. Chun, and R.~Mandal, ``{Implications of right-handed
  neutrinos in $B-L$ extended standard model with scalar dark matter},''
  \href{http://dx.doi.org/10.1103/PhysRevD.97.015001}{{\em Phys. Rev. D}
  {\bfseries 97} no.~1, (2018) 015001},
  \href{http://arxiv.org/abs/1707.00874}{{\ttfamily arXiv:1707.00874
  [hep-ph]}}.

\bibitem{Das:2019pua}
A.~Das, S.~Goswami, K.~N. Vishnudath, and T.~Nomura, ``{Constraining a general
  U(1)$^\prime$ inverse seesaw model from vacuum stability, dark matter and
  collider},'' \href{http://dx.doi.org/10.1103/PhysRevD.101.055026}{{\em Phys.
  Rev. D} {\bfseries 101} no.~5, (2020) 055026},
  \href{http://arxiv.org/abs/1905.00201}{{\ttfamily arXiv:1905.00201
  [hep-ph]}}.

\bibitem{Minkowski:1977sc}
P.~Minkowski, ``{$\mu \to e\gamma$ at a Rate of One Out of $10^{9}$ Muon
  Decays?},'' \href{http://dx.doi.org/10.1016/0370-2693(77)90435-X}{{\em Phys.
  Lett. B} {\bfseries 67} (1977) 421--428}.

\bibitem{Schechter:1980gr}
J.~Schechter and J.~W.~F. Valle, ``{Neutrino Masses in SU(2) x U(1)
  Theories},'' \href{http://dx.doi.org/10.1103/PhysRevD.22.2227}{{\em Phys.
  Rev. D} {\bfseries 22} (1980) 2227}.

\bibitem{Mohapatra:1979ia}
R.~N. Mohapatra and G.~Senjanovic, ``{Neutrino Mass and Spontaneous Parity
  Nonconservation},'' \href{http://dx.doi.org/10.1103/PhysRevLett.44.912}{{\em
  Phys. Rev. Lett.} {\bfseries 44} (1980) 912}.

\bibitem{Schechter:1981cv}
J.~Schechter and J.~W.~F. Valle, ``{Neutrino Decay and Spontaneous Violation of
  Lepton Number},'' \href{http://dx.doi.org/10.1103/PhysRevD.25.774}{{\em Phys.
  Rev. D} {\bfseries 25} (1982) 774}.

\bibitem{Ma:2006km}
E.~Ma, ``{Verifiable radiative seesaw mechanism of neutrino mass and dark
  matter},'' \href{http://dx.doi.org/10.1103/PhysRevD.73.077301}{{\em Phys.
  Rev. D} {\bfseries 73} (2006) 077301},
  \href{http://arxiv.org/abs/hep-ph/0601225}{{\ttfamily arXiv:hep-ph/0601225}}.

\bibitem{Hirsch:2013ola}
M.~Hirsch, R.~A. Lineros, S.~Morisi, J.~Palacio, N.~Rojas, and J.~W.~F. Valle,
  ``{WIMP dark matter as radiative neutrino mass messenger},''
  \href{http://dx.doi.org/10.1007/JHEP10(2013)149}{{\em JHEP} {\bfseries 10}
  (2013) 149}, \href{http://arxiv.org/abs/1307.8134}{{\ttfamily arXiv:1307.8134
  [hep-ph]}}.

\bibitem{Merle:2016scw}
A.~Merle, M.~Platscher, N.~Rojas, J.~W.~F. Valle, and A.~Vicente,
  ``{Consistency of WIMP Dark Matter as radiative neutrino mass messenger},''
  \href{http://dx.doi.org/10.1007/JHEP07(2016)013}{{\em JHEP} {\bfseries 07}
  (2016) 013}, \href{http://arxiv.org/abs/1603.05685}{{\ttfamily
  arXiv:1603.05685 [hep-ph]}}.

\bibitem{Avila:2019hhv}
I.~M. \'Avila, V.~De~Romeri, L.~Duarte, and J.~W.~F. Valle, ``{Phenomenology of
  scotogenic scalar dark matter},''
  \href{http://dx.doi.org/10.1140/epjc/s10052-020-08480-z}{{\em Eur. Phys. J.
  C} {\bfseries 80} no.~10, (2020) 908},
  \href{http://arxiv.org/abs/1910.08422}{{\ttfamily arXiv:1910.08422
  [hep-ph]}}.

\bibitem{Mandal:2021yph}
S.~Mandal, R.~Srivastava, and J.~W.~F. Valle, ``{The simplest scoto-seesaw
  model: WIMP dark matter phenomenology and Higgs vacuum stability},''
  \href{http://dx.doi.org/10.1016/j.physletb.2021.136458}{{\em Phys. Lett. B}
  {\bfseries 819} (2021) 136458},
  \href{http://arxiv.org/abs/2104.13401}{{\ttfamily arXiv:2104.13401
  [hep-ph]}}.

\bibitem{Mandal:2019oth}
S.~Mandal, N.~Rojas, R.~Srivastava, and J.~W.~F. Valle, ``{Dark matter as the
  origin of neutrino mass in the inverse seesaw mechanism},''
  \href{http://dx.doi.org/10.1016/j.physletb.2021.136609}{{\em Phys. Lett. B}
  {\bfseries 821} (2021) 136609},
  \href{http://arxiv.org/abs/1907.07728}{{\ttfamily arXiv:1907.07728
  [hep-ph]}}.

\bibitem{Das:2022oyx}
A.~Das, S.~Gola, S.~Mandal, and N.~Sinha, ``{Two-component scalar and fermionic
  dark matter candidates in a generic U(1)X model},''
  \href{http://dx.doi.org/10.1016/j.physletb.2022.137117}{{\em Phys. Lett. B}
  {\bfseries 829} (2022) 137117},
  \href{http://arxiv.org/abs/2202.01443}{{\ttfamily arXiv:2202.01443
  [hep-ph]}}.

\bibitem{Darvishi_2021}
N.~Darvishi, M.~Masouminia, and A.~Pilaftsis, ``Maximally symmetric
  three-higgs-doublet model,''
  \href{http://dx.doi.org/10.1103/physrevd.104.115017}{{\em Physical Review D}
  {\bfseries 104} no.~11, (Dec, 2021) }.
  \url{https://doi.org/10.1103%2Fphysrevd.104.115017}.

\bibitem{Robens_2020}
T.~Robens, T.~Stefaniak, and J.~Wittbrodt, ``Two-real-scalar-singlet extension
  of the {SM}: {LHC} phenomenology and benchmark scenarios,''
  \href{http://dx.doi.org/10.1140/epjc/s10052-020-7655-x}{{\em The European
  Physical Journal C} {\bfseries 80} no.~2, (Feb, 2020) }.
  \url{https://doi.org/10.1140%2Fepjc%2Fs10052-020-7655-x}.

\bibitem{Kannike:2012pe}
K.~Kannike, ``{Vacuum Stability Conditions From Copositivity Criteria},''
  \href{http://dx.doi.org/10.1140/epjc/s10052-012-2093-z}{{\em Eur. Phys. J. C}
  {\bfseries 72} (2012) 2093}, \href{http://arxiv.org/abs/1205.3781}{{\ttfamily
  arXiv:1205.3781 [hep-ph]}}.

\bibitem{Djouadi_2008}
A.~Djouadi, ``The anatomy of electroweak symmetry breaking,''
  \href{http://dx.doi.org/10.1016/j.physrep.2007.10.004}{{\em Physics Reports}
  {\bfseries 457} no.~1-4, (Feb, 2008) 1--216}.
  \url{https://doi.org/10.1016%2Fj.physrep.2007.10.004}.

\bibitem{ATLAS:2020kdi}
{\bfseries ATLAS} Collaboration, ``{Combination of searches for invisible Higgs
  boson decays with the ATLAS experiment},''.

\bibitem{ATLAS:2019cid}
{\bfseries ATLAS} Collaboration, M.~Aaboud {\em et~al.}, ``{Combination of
  searches for invisible Higgs boson decays with the ATLAS experiment},''
  \href{http://dx.doi.org/10.1103/PhysRevLett.122.231801}{{\em Phys. Rev.
  Lett.} {\bfseries 122} no.~23, (2019) 231801},
  \href{http://arxiv.org/abs/1904.05105}{{\ttfamily arXiv:1904.05105
  [hep-ex]}}.

\bibitem{CMS:2018yfx}
{\bfseries CMS} Collaboration, A.~M. Sirunyan {\em et~al.}, ``{Search for
  invisible decays of a Higgs boson produced through vector boson fusion in
  proton-proton collisions at $\sqrt{s} =$ 13 TeV},''
  \href{http://dx.doi.org/10.1016/j.physletb.2019.04.025}{{\em Phys. Lett. B}
  {\bfseries 793} (2019) 520--551},
  \href{http://arxiv.org/abs/1809.05937}{{\ttfamily arXiv:1809.05937
  [hep-ex]}}.

\bibitem{Planck:2018vyg}
{\bfseries Planck} Collaboration, N.~Aghanim {\em et~al.}, ``{Planck 2018
  results. VI. Cosmological parameters},''
  \href{http://dx.doi.org/10.1051/0004-6361/201833910}{{\em Astron. Astrophys.}
  {\bfseries 641} (2020) A6}, \href{http://arxiv.org/abs/1807.06209}{{\ttfamily
  arXiv:1807.06209 [astro-ph.CO]}}.

\bibitem{Baring_2016}
M.~G. Baring, T.~Ghosh, F.~S. Queiroz, and K.~Sinha, ``New limits on the dark
  matter lifetime from dwarf spheroidal galaxies using fermi-{LAT},''
  \href{http://dx.doi.org/10.1103/physrevd.93.103009}{{\em Physical Review D}
  {\bfseries 93} no.~10, (May, 2016) }.
  \url{https://doi.org/10.1103%2Fphysrevd.93.103009}.

\bibitem{Belanger:2018ccd}
G.~B\'elanger, F.~Boudjema, A.~Goudelis, A.~Pukhov, and B.~Zaldivar,
  ``{micrOMEGAs5.0 : Freeze-in},''
  \href{http://dx.doi.org/10.1016/j.cpc.2018.04.027}{{\em Comput. Phys.
  Commun.} {\bfseries 231} (2018) 173--186},
  \href{http://arxiv.org/abs/1801.03509}{{\ttfamily arXiv:1801.03509
  [hep-ph]}}.

\bibitem{Aprile:2018dbl}
{\bfseries XENON} Collaboration, E.~Aprile {\em et~al.}, ``{Dark Matter Search
  Results from a One Ton-Year Exposure of XENON1T},''
  \href{http://dx.doi.org/10.1103/PhysRevLett.121.111302}{{\em Phys. Rev.
  Lett.} {\bfseries 121} no.~11, (2018) 111302},
  \href{http://arxiv.org/abs/1805.12562}{{\ttfamily arXiv:1805.12562
  [astro-ph.CO]}}.

\bibitem{Akerib:2018lyp}
{\bfseries LUX-ZEPLIN} Collaboration, D.~S. Akerib {\em et~al.}, ``{Projected
  WIMP sensitivity of the LUX-ZEPLIN dark matter experiment},''
  \href{http://dx.doi.org/10.1103/PhysRevD.101.052002}{{\em Phys. Rev. D}
  {\bfseries 101} no.~5, (2020) 052002},
  \href{http://arxiv.org/abs/1802.06039}{{\ttfamily arXiv:1802.06039
  [astro-ph.IM]}}.

\bibitem{O_Hare_2021}
C.~A.~J. O'Hare, ``New definition of the neutrino floor for direct dark matter
  searches,'' \href{http://dx.doi.org/10.1103/physrevlett.127.251802}{{\em
  Physical Review Letters} {\bfseries 127} no.~25, (Dec, 2021) }.
  \url{https://doi.org/10.1103%2Fphysrevlett.127.251802}.

\bibitem{Ishiwata_2018}
K.~Ishiwata and T.~Toma, ``Probing pseudo nambu-goldstone boson dark matter at
  loop level,'' \href{http://dx.doi.org/10.1007/jhep12(2018)089}{{\em Journal
  of High Energy Physics} {\bfseries 2018} no.~12, (Dec, 2018) }.
  \url{https://doi.org/10.1007%2Fjhep12%282018%29089}.

\bibitem{deBlas:2018mhx}
J.~de~Blas {\em et~al.}, ``{The CLIC Potential for New Physics},''
  \href{http://arxiv.org/abs/1812.02093}{{\ttfamily arXiv:1812.02093
  [hep-ph]}}.

\bibitem{LEPWorkingGroupforHiggsbosonsearches:2003ing}
{\bfseries LEP Working Group for Higgs boson searches, ALEPH, DELPHI, L3, OPAL}
  Collaboration, R.~Barate {\em et~al.}, ``{Search for the standard model Higgs
  boson at LEP},'' \href{http://dx.doi.org/10.1016/S0370-2693(03)00614-2}{{\em
  Phys. Lett. B} {\bfseries 565} (2003) 61--75},
  \href{http://arxiv.org/abs/hep-ex/0306033}{{\ttfamily arXiv:hep-ex/0306033}}.

\bibitem{Wang:2020lkq}
Y.~Wang, M.~Berggren, and J.~List, ``{ILD Benchmark: Search for Extra Scalars
  Produced in Association with a $Z$ boson at $\sqrt{s}=500$ GeV},''
  \href{http://arxiv.org/abs/2005.06265}{{\ttfamily arXiv:2005.06265
  [hep-ex]}}.

\bibitem{ATLAS:2020qdt}
{\bfseries ATLAS} Collaboration, ``{A combination of measurements of Higgs
  boson production and decay using up to $139\,\text{fb}^{-1}$ of $pp$
  collision data at $\sqrt{s}=$ 13 TeV collected with the ATLAS experiment},''
  8, 2020.

\bibitem{Buttazzo:2018qqp}
D.~Buttazzo, D.~Redigolo, F.~Sala, and A.~Tesi, ``{Fusing Vectors into Scalars
  at High Energy Lepton Colliders},''
  \href{http://dx.doi.org/10.1007/JHEP11(2018)144}{{\em JHEP} {\bfseries 11}
  (2018) 144}, \href{http://arxiv.org/abs/1807.04743}{{\ttfamily
  arXiv:1807.04743 [hep-ph]}}.

\end{thebibliography}\endgroup
\end{document}